\documentclass[aps,prb,reprint,longbibliography]{revtex4-2}
\usepackage{graphicx}
\usepackage{mathrsfs}
\usepackage{bm}
\usepackage{amssymb}
\usepackage{amsmath}
\usepackage{textcomp}
\usepackage{url}
\usepackage{dsfont}	
\usepackage{braket}
\usepackage[dvipsnames]{xcolor} 
\usepackage[colorlinks=true,linkcolor=magenta,citecolor=blue]{hyperref}
\usepackage{empheq}
\usepackage{multirow}
\begin{document}

\preprint{APS/123-QED}

\title{Symmetry-Enforced Topology and Van Hove Saddle Points in Hexagonal $\mathrm{TiAs}$}

\author{Azizur Rahman$^1$}
\thanks{These authors contributed equally to this work.}
\author{Shashi B Mishra$^2$}
\thanks{These authors contributed equally to this work.}
\author{Sheraz khan$^1$}
\author{Khushnood Shah$^1$}
\author{Xun Ma$^3$}
\author{Xiaoyang Lin$^1$}
\author{Tomasz Sobol$^4$}
\author{Lei Zhang$^5$}
\author{Weisheng Zhao$^1$}
\author{Min Zhang$^6$}
\email{mzhang@ahu.edu.cn}
\author{Zahir Muhammad$^1$}
\email{zahir@buaa.edu.cn}

\affiliation{$^1$National Key Laboratory of Spintronics, Hangzhou International Innovation Institute, 
Beihang University, Hangzhou 311115, P.R. China}
\affiliation{$^2$Department of Electrical and Computer Engineering, University of Maryland, College Park, MD 20742, USA}
\affiliation{$^3$School of Nuclear Science and Technology, University of Science and Technology of China, Hefei, Anhui, China.}
\affiliation{$^4$SOLARIS National Synchrotron Radiation Centre, Jagiellonian University, Krakow, Poland}
\affiliation{$^5$Anhui Key Laboratory of Low-Energy Quantum Materials and Devices,
High Magnetic Field Laboratory, Hefei Institutes of Physical Science,
Chinese Academy of Sciences, Hefei 230031, China}
\affiliation{$^6$Institutes of Physical Science and Information Technology, Anhui University, Hefei 230601, China}

\begin{abstract}

We report the first successful growth of stoichiometric hexagonal TiAs single crystals (space group $P6_3/mmc$) using chemical vapor transport and investigate their vibrational, electronic, topological, and magnetotransport properties. Polarized Raman spectroscopy identifies all four Raman-active phonon modes ($A_{1g} + E_{1g} + 2E_{2g}$), consistent with the hexagonal crystal symmetry. First-principles calculations reveal nontrivial band topology characterized by a symmetry indicator $\mathbb{Z}_4 = 2$ and a strong topological index $\nu_0 = 1$, together with symmetry-protected fourfold band degeneracies at the nonsymmorphic Brillouin-zone boundary. The Fermi surface exhibits an orbital-dependent crossover from quasi-two-dimensional cylindrical sheets to a corrugated three-dimensional structure along $k_z$. A Van Hove saddle point approximately 0.5 eV below the Fermi level along $\Gamma\text{--}K$ is associated with a neck-disrupting Lifshitz transition and enhanced density of states. Magnetotransport measurements reveal a pronounced twofold angular magnetoresistance, consistent with anisotropic Fermi-surface geometry and open cyclotron orbits. Hall measurements indicate uncompensated, hole-dominated transport, accounting for the absence of extreme magnetoresistance and contributing to the field-orientation-dependent Hall response. These results establish hexagonal TiAs as a platform for exploring the interplay of nonsymmorphic band topology, Van Hove singularities, and anisotropic electronic transport in transition-metal monopnictides.

\end{abstract}

\maketitle
\section{INTRODUCTION}

The discovery of topological electronic phases has reshaped modern condensed matter physics, establishing band topology alongside spontaneous symmetry breaking as a cornerstone for classifying quantum states of matter \cite{hasan2010colloquium,qi2011topological}. Over the past decade, the conceptual frontier has expanded from gapped topological insulators to gapless topological semimetals, where conduction and valence bands touch at discrete nodal points or along continuous nodal lines in momentum space \cite{armitage2018weyl,lv2021experimental}. Early prototypes, such as the three-dimensional (3D) Dirac semimetals $\mathrm{Cd_3As_2}$ \cite{neupane2014observation,borisenko2014experimental} and $\mathrm{Na_3Bi}$ \cite{liu2014discovery}, as well as the non-centrosymmetric $\mathrm{TaAs}$ family of Weyl semimetals \cite{lv2015experimental,xu2015discovery}, have revealed an array of exotic quantum phenomena, including ultrahigh carrier mobilities, extremely large non-saturating magnetoresistance (XMR), chiral anomaly-induced negative longitudinal magnetoresistance, and topologically protected surface Fermi arcs \cite{shekhar2015extremely,ali2014large,xiong2015evidence}.

However, in many early semimetallic systems, the gapless band crossings arise from accidental band inversions governed by chemical stoichiometry and fine-tuned orbital hybridization \cite{armitage2018weyl,yang2014dirac}. Such accidental nodes are inherently vulnerable to lattice distortions, strain, or chemical perturbations that can open trivial hybridization gaps. To overcome this limitation, recent theoretical developments grounded in space-group representation theory and topological quantum chemistry (TQC) have shifted significant attention toward \emph{symmetry-enforced} topological semimetals \cite{bradlyn2017topological,po2017symmetry,vergniory2019complete}. In these materials, the filling constraints of the space group preclude the realization of an atomic insulator, guaranteeing gapless band degeneracies at high-symmetry points, along lines, or over entire boundary planes of the Brillouin zone solely by virtue of the crystal symmetries \cite{young2012dirac,tang2019comprehensive}. In particular, nonsymmorphic space groups featuring fractional lattice translations, such as screw rotations and glide mirrors—enforce multi-band connectivity and higher-dimensional band degeneracies that are robust against any symmetry-preserving perturbation \cite{wieder2018wallpaper,bradlyn2017topological}.

Among nonsymmorphic candidates, transition-metal monopnictides of the hexagonal $\mathrm{NiAs}$ structural family (space group $P6_3/mmc$, No.~194) offer an exceptional, yet experimentally under-explored, platform. In this centrosymmetric lattice, the coexistence of spatial inversion ($\mathcal{P}$), time-reversal ($\mathcal{T}$), and a nonsymmorphic sixfold screw axis $\{C_{6z}|\boldsymbol{\tau}\}$ with $\boldsymbol{\tau} = (0, 0, \frac{1}{2})$ forces two Kramers doublets to merge into fourfold-degenerate Dirac states across the entire $k_z = \pi/c$ zone-boundary plane \cite{young2012dirac,bradlyn2017topological}. Although theoretical symmetry-indicator databases predict nontrivial semimetallic states across titanium- and zirconium-based monopnictides \cite{vergniory2019complete,zhang2019catalogue}, synthesizing high-purity, defect-free single crystals of titanium monoarsenide ($\mathrm{TiAs}$) has remained challenging due to the elevated vapor pressure of arsenic and the stability of competing multiphase binaries. Consequently, a direct experimental determination of the bulk electronic structure, $k_z$-dependent band dispersion, and macroscopic carrier dynamics in stoichiometric $\mathrm{TiAs}$ is notably lacking.

In this work, we report the single-crystal growth of stoichiometric $\mathrm{TiAs}$ via iodine-assisted chemical vapor transport and provide a unified picture of its nonsymmorphic band topology and carrier dynamics. Combining photon-energy-dependent ARPES tomography with relativistic first-principles calculations, we map the three-dimensional low-energy electronic structure and find a nontrivial parity indicator $\mathbb{Z}_4=2$ with symmetry-enforced band connectivity, establishing $\mathrm{TiAs}$ as a symmetry-enforced semimetal. Electrical magnetotransport measurements reveal a pronounced twofold angular anisotropy driven by directional open-orbit cyclotron paths, alongside multi-band Hall dynamics reflecting uncompensated electron and hole conduction. Together, these results establish $\mathrm{TiAs}$ as a model hexagonal monopnictide for exploring symmetry-enforced band crossings and directional quantum transport in nonsymmorphic semimetals.

\section{Single-Crystal Growth}

Polycrystalline precursors and high-quality single crystals of $\mathrm{TiAs}$ were synthesized via solid-state reaction followed by chemical vapor transport (CVT). High-purity $\mathrm{Ti}$ powder ($\ge 99.99\%$) and $\mathrm{As}$ granules ($\ge 99.999\%$) were weighed in an equimolar ratio ($\mathrm{Ti}:\mathrm{As} = 1:1$) with a total batch mass of $1.0\text{--}2.0\text{ g}$. The starting elements were thoroughly homogenized and transferred into a quartz ampoule (inner diameter $10\text{--}12\text{ mm}$, length $12\text{--}15\text{ cm}$). The ampoule was evacuated to a base pressure below $5 \times 10^{-3}\text{ Pa}$, flame-sealed, and placed in a horizontal tube furnace. The reaction mixture was heated to $750\,^{\circ}\mathrm{C}$ at a ramp rate of $50\,^{\circ}\mathrm{C}/\mathrm{h}$ and held at this temperature for $48\text{ h}$ to ensure complete reaction into the binary phase. After naturally furnace-cooling to room temperature, the quartz ampoule was opened, and the synthesized polycrystalline $\mathrm{TiAs}$ ingots were lightly pulverized into a homogeneous precursor powder.

\begin{figure*}[htbp]
	\includegraphics[width=0.9\linewidth]{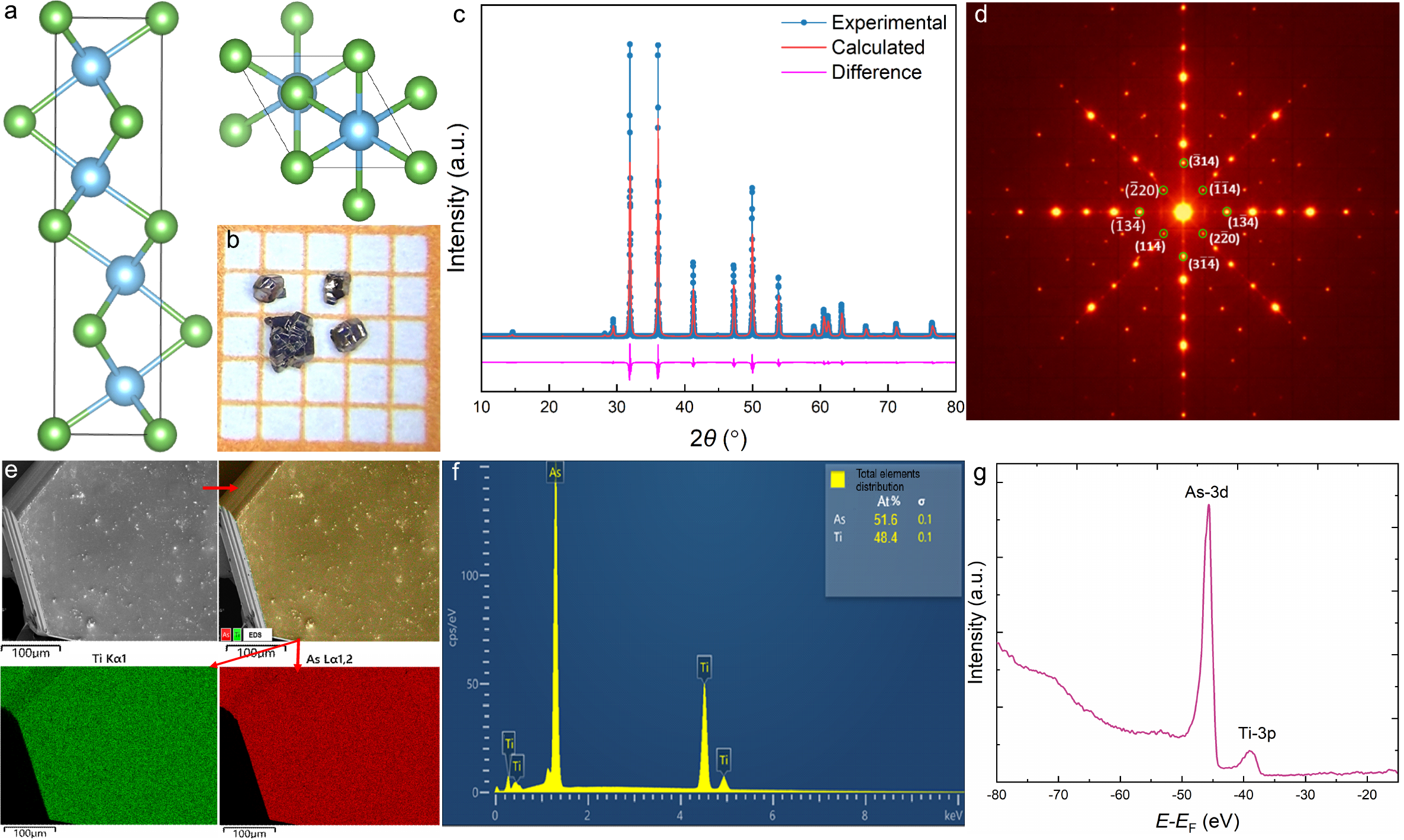}
	\caption{\textbf{Structural and chemical characterization of $\mathrm{TiAs}$.} 
(a) Hexagonal $\mathrm{TiAs}$-type unit cell (space group $P6_3/mmc$; $\mathrm{Ti}$ light blue, $\mathrm{As}$ green) and, right, octahedral coordination of $\mathrm{Ti}$ by six $\mathrm{As}$ atoms. 
(b) As-grown single crystals exhibiting well-developed hexagonal facets on a millimeter grid. 
(c) Powder XRD pattern and Rietveld refinement: experimental data (blue circles), calculated profile (red line), and difference curve (magenta line). 
(d) Single-crystal X-ray Laue back-reflection pattern displaying sharp, indexed diffraction spots with hexagonal symmetry. 
(e) Backscattered-electron SEM micrograph of a flat crystal facet alongside elemental EDS maps for $\mathrm{Ti}$ ($K\alpha_1$) and $\mathrm{As}$ ($L\alpha_{1,2}$). 
(f) Energy-dispersive X-ray spectrum (EDS) confirming near-equiatomic stoichiometry (inset: extracted atomic percentages). 
(g) Core-level photoemission spectrum resolving the $\mathrm{As}\ 3d$ and $\mathrm{Ti}\ 3p$ core states referenced to $E_{\mathrm{F}}$.}
	\label{fig:1}
\end{figure*}

For the chemical vapor transport growth of single crystals, $0.5\text{--}1.0\text{ g}$ of the polycrystalline $\mathrm{TiAs}$ precursor was blended with resublimed iodine ($\mathrm{I}_2$) serving as the transport agent, with the transport agent concentration maintained at $5\text{--}10\text{ mg}/\mathrm{cm}^3$ relative to the internal ampoule volume. The mixture was loaded into a clean quartz tube, with the charge positioned at the source end.

\section{Results and discussion}
The crystal structure of TiAs, schematically illustrated in Fig.~\ref{fig:1}a, is the hexagonal TiAs type, a NiAs-derived structure in the centrosymmetric space group $P6_3/mmc$ (No.~194) with four formula units per cell \cite{bachmayer1955,kjekshus1964}. The cell is doubled along $c$ relative to NiAs ($c/a = 3.32$, Table~\ref{tab:table1}) because the As sublattice adopts the double-hexagonal $ABAC$ stacking rather than the simple hcp sequence of the parent structure. This renders the two As sites inequivalent: As at $(0,0,0)$ is octahedrally coordinated by six Ti at $2.538\ \text{\AA}$, whereas As at $(\tfrac{2}{3},\tfrac{1}{3},\tfrac{1}{4})$ sits at the center of a trigonal prism of six Ti at $2.649\ \text{\AA}$; every Ti, in turn, is octahedrally coordinated by six As. The resulting TiAs$_6$ octahedra share faces in pairs along the $c$-axis, consecutive pairs being joined by edge sharing, which yields short Ti--Ti distances of $3.215\ \text{\AA}$ that promote a hybridization of covalent-metallic Ti--As bonding alongside direct metal--metal interactions. The as-grown single crystals exhibit a bright metallic luster with highly symmetric, well-developed hexagonal facets (Fig.~\ref{fig:1}b), with typical lateral dimensions ranging from several hundred micrometers up to a few millimeters.

Phase purity and structural parameters were comprehensively evaluated via powder X-ray diffraction (XRD) on crushed single crystals. As demonstrated in Fig.~\ref{fig:1}c, the experimental diffraction profile (blue) exhibits perfect match with the calculated profile (red) simulated from the hexagonal $P6_3/mmc$ phase. All observed reflections were unambiguously indexed to the TiAs phase with no discernible secondary phases or elemental impurity peaks. Rietveld refinement performed using MAUD software \cite{maud_ref} yielded high-quality convergence reliability factors ($R_{\text{wp}} = 8.2\%$, $\chi^2 = 1.85$). The refinement model incorporated pseudo-Voigt peak shapes, axial divergence corrections, a Chebyshev polynomial background, and a March--Dollase preferred orientation model \cite{march_dollase} along the $[0001]$ direction to account for the plate-like crystal morphology. The refined lattice parameters show remarkable agreement with both experimental literature values (ICSD~44981) \cite{bachmayer1955} and computationally optimized structures from the Materials Project database (mp-1822) \cite{mp1822}, displaying minimal volume deviation (Table~\ref{tab:table1}). To verify single crystallinity and determine the crystallographic orientation of the cleaved facets, single-crystal X-ray Laue back-reflection measurements were conducted [Fig.~\ref{fig:1}(d)]. The pattern exhibits sharp, well-defined diffraction spots arranged in clear sixfold symmetry without streaking or splitting, confirming high crystalline quality, long-range order, and the absence of twinning or secondary grain boundaries. The indexing of low-index Bragg reflections, such as $(3\bar{1}4)$ and $(\bar{1}34)$, demonstrates that the broad, flat cleavage plane corresponds to the crystallographic basal plane perpendicular to the $c$ axis.

The elemental composition and spatial homogeneity of the as-grown crystals were investigated using scanning electron microscopy coupled with energy-dispersive X-ray spectroscopy (SEM-EDS). Fig.~\ref{fig:1}e displays a representative backscattered electron (BSE) image of a polished crystal surface alongside corresponding elemental distribution maps. Both Ti (green) and As (red) are uniformly distributed across the crystal matrix without detectable elemental segregation, clustering, or secondary-phase inclusions. Quantitative EDS analysis (Fig.~\ref{fig:1}f) yields an empirical composition of $48.4 \pm 0.1\text{ at}\%\text{ Ti}$ and $51.6 \pm 0.1\text{ at}\%\text{ As}$. This indicates a slight arsenic excess relative to ideal $1:1$ stoichiometry, a characteristic feature frequently documented in transition-metal pnictides \cite{kjekshus1964} that typically stems from local surface termination effects or a narrow off-stoichiometric homogeneity range inherent to the phase growth conditions.

\begin{table}[htbp]
	\centering
	\caption{\textbf{Refined structural parameters of TiAs.} Rietveld refinement of this work (MAUD)
		compared with DFT-GGA calculations (Materials Project mp-1822)~\cite{mp1822} and experimental literature
		data (ICSD 44981)~\cite{bachmayer1955}. $\Delta$ is the relative difference between this work and mp-1822;
		$R_{\mathrm{wp}}$ and $\chi^2$ are the weighted profile residual and the goodness of fit.}
	\label{tab:table1}
	\renewcommand{\arraystretch}{1.15}
	\begin{tabular*}{\columnwidth}{@{\extracolsep{\fill}}lcccc}
		\hline\hline
		Parameter & This work & mp-1822 & ICSD 44981 & $\Delta$ (\%) \\
		\hline
		$a$ (\AA) & 3.642 & 3.649 & 3.64 & $-0.19$ \\
		$c$ (\AA) & 12.092 & 12.118 & $\sim$12.10--12.16 & $-0.21$ \\
		$c/a$ & 3.320 & 3.320 & $\sim$3.325 & $0.00$ \\
		$V$ (\AA$^3$) & 139.05 & 139.74 & $\sim$139.0--140.0 & $-0.50$ \\
		$R_{\mathrm{wp}}$ (\%) & 8.2 & --- & --- & --- \\
		$\chi^2$ & 1.85 & --- & --- & --- \\
		\hline\hline
	\end{tabular*}
\end{table}

To probe the surface chemical states and valence characteristics, core-level photoemission spectroscopy was performed. As shown in Fig.~\ref{fig:1}g, the core-level spectrum referenced to the Fermi level ($E - E_F = 0\text{ eV}$) clearly resolves the prominent $\text{As } 3d$ doublet envelope and the shallow $\text{Ti } 3p$ core level, confirming the expected chemical bonding environment of the TiAs matrix. The observed binding energies exhibit minor shifts relative to standard elemental reference values \cite{moulder1992xps}, consistent with charge transfer during Ti--As bond formation alongside minimal surface oxidation from ambient handling. The sharp, well-resolved peak profiles and intensity ratios corroborate the structural and chemical integrity of the synthesized single crystals.

\begin{figure*}[htbp]
	\includegraphics[width=0.8\linewidth]{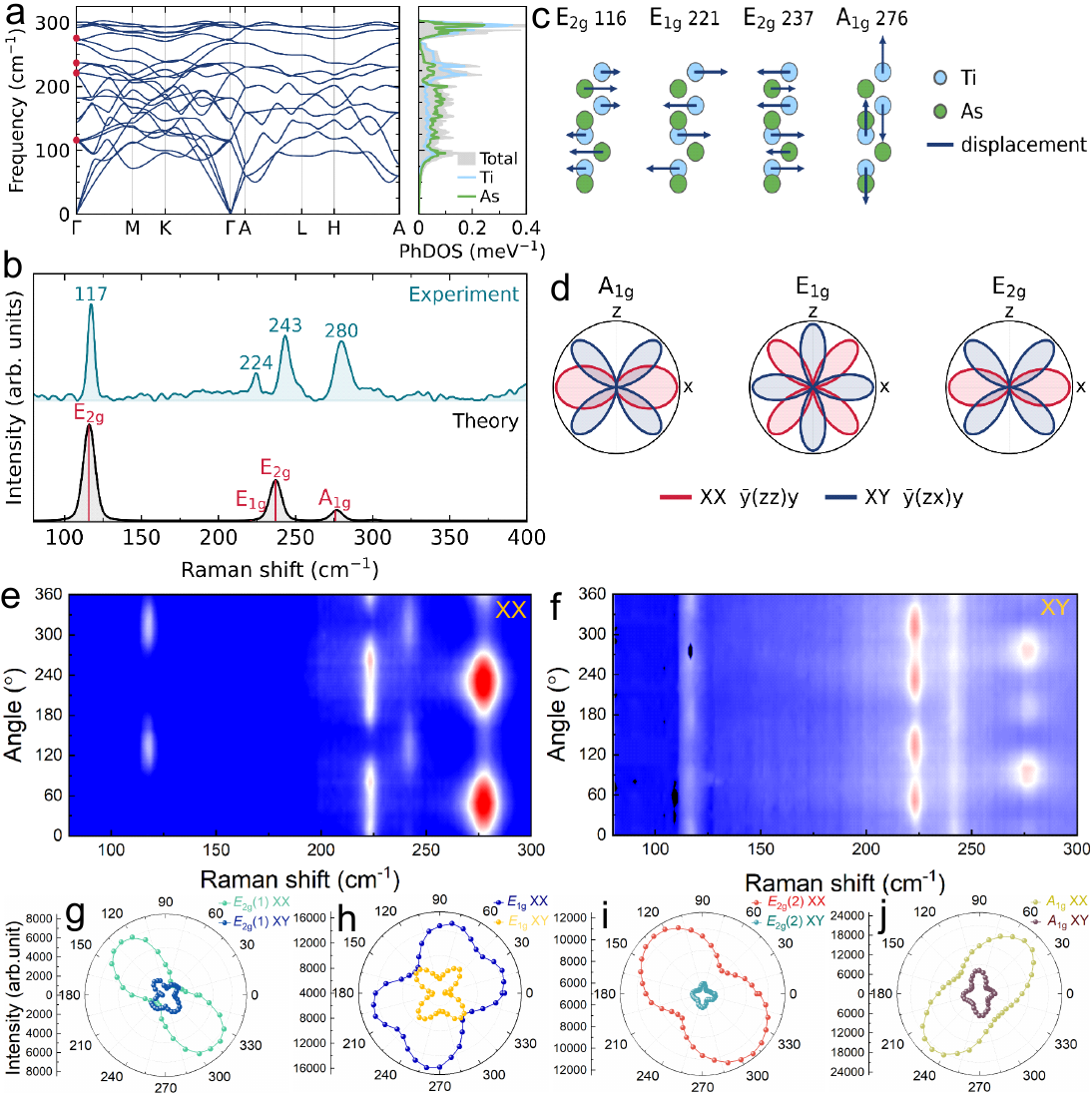}
	\caption{\textbf{Phonon dispersion and vibrational analysis.} 
		(a) Phonon dispersion along the high-symmetry path with the phonon density of states (PhDOS): total (grey), $\mathrm{Ti}$ (light blue) and $\mathrm{As}$ (green). Red circles mark the four Raman-active $\Gamma$ modes. 
		(b) Displacement eigenvectors of those modes in the $a\text{--}c$ plane; arrow length scales with amplitude, atoms colored as in Fig.~\ref{fig:1}a. Both sublattices move in $E_{2g}(1)$, whereas $E_{1g}$, $E_{2g}(2)$ and $A_{1g}$ are dominated by $\mathrm{Ti}$. 
		(c) Resonant Raman spectrum at $E_L = 2.54\text{ eV}$, Gaussian-broadened by $8\text{ cm}^{-1}$ (black), against the measured unpolarized $300\text{ K}$ spectrum (teal). Red sticks give the calculated mode positions and intensities, labeled by irreducible representation. 
		(d) Angular dependence $I(\theta)$ of each irreducible representation in the collinear ($XX$, $\bar{y}(zz)y$; red) and cross-linear ($XY$, $\bar{y}(zx)y$; navy) backscattering geometries, giving the expected twofold and fourfold patterns. (e,~f) Intensity maps versus azimuthal angle ($0^\circ\text{--}360^\circ$) in the (a) collinear ($XX$, $\bar{y}(zz)y$) and (g) cross-linear ($XY$, $\bar{y}(zx)y$) backscattering geometries, at $300\text{ K}$ under $532\text{ nm}$ excitation ($500\ \mu\mathrm{W}$). (g) $E_{2g}(1)$, $117\text{ cm}^{-1}$; (h) $E_{1g}$, $224\text{ cm}^{-1}$; (i) $E_{2g}(2)$, $243\text{ cm}^{-1}$; (j) $A_{1g}$, $280\text{ cm}^{-1}$. Markers are experiment, curves least-squares fits to the Raman tensor selection rules.}
	\label{fig:2}
\end{figure*}

To establish a microscopic baseline for the vibrational properties of TiAs, first-principles lattice dynamics calculations were performed using density functional perturbation theory (DFPT)~\cite{Baroni2001}. For centrosymmetric hexagonal TiAs [$D_{6h}^{4}$, No.~194], site-symmetry and factor group analysis at $\Gamma$  decompose the total optical phonon representation into $\Gamma_{\mathrm{opt}} = A_{1g} + E_{1g} + 2E_{2g} + 2A_{2u} + 2E_{1u} + 2B_{1g} + 2B_{2u} + 2E_{2u}$, yielding exactly four first-order Raman-active modes ($A_{1g} + E_{1g} + 2E_{2g}$), four infrared-active modes ($2A_{2u} + 2E_{1u}$), and six optically silent optical branches~\cite{fateley1971infrared,kroumova2003bilbao}. The calculated phonon dispersion [Fig.~\ref{fig:2}a] contains no imaginary frequencies, confirming the dynamical stability of the synthesized $P6_3/mmc$ phase. The atom-projected phonon density of states reveals substantial Ti--As mixing: As contributes $63\%$ of the states below $180\text{ cm}^{-1}$, whereas Ti contributes $64\%$ above $200\text{ cm}^{-1}$. This overlap reflects the relatively small mass ratio $m_{\mathrm{As}}/m_{\mathrm{Ti}} = 1.57$, in contrast to $\mathrm{TaAs}$ ($m_{\mathrm{Ta}}/m_{\mathrm{As}} = 2.4$), where the phonon manifolds are more clearly separated~\cite{Liu2025taas,Mishra2026taas}.
The Raman-mode eigenvectors [Fig.~\ref{fig:2}b] reflect the same behavior. The lowest-frequency $E_{2g}(1)$ mode ($116\text{ cm}^{-1}$) is a mixed in-plane Ti--As shear with nearly equal displacement contributions ($48\%$ Ti and $52\%$ As). By contrast, the in-plane $E_{1g}$ ($221\text{ cm}^{-1}$) and $E_{2g}(2)$ ($237\text{ cm}^{-1}$) modes are predominantly Ti-derived ($100\%$ and $87\%$, respectively), while the $A_{1g}$ mode ($274\text{ cm}^{-1}$) is an out-of-plane Ti breathing vibration along the $c$ axis.

To quantitatively describe the angular dependence of the scattered intensities, the Raman-active modes are modeled using the corresponding $3 \times 3$ Raman tensors ($\mathcal{R}$) under $D_{6h}$ point-group symmetry \cite{loudon1964raman}. In the Cartesian reference frame where $z \parallel c$ and the $x, y$ axes lie within the hexagonal basal plane, the non-zero independent tensor components take the form \cite{loudon1964raman}:
\begin{equation}
	\setlength{\arraycolsep}{3pt}
	\begin{aligned}
		\mathcal{R}(A_{1g}) &= \begin{pmatrix} a & 0 & 0 \\ 0 & a & 0 \\ 0 & 0 & b \end{pmatrix}, \\
		\mathcal{R}(E_{1g}) &= \left\{ \begin{pmatrix} 0 & 0 & -c \\ 0 & 0 & 0 \\ -c & 0 & 0 \end{pmatrix}, \, \begin{pmatrix} 0 & 0 & 0 \\ 0 & 0 & c \\ 0 & c & 0 \end{pmatrix} \right\}, \\
		\mathcal{R}(E_{2g}) &= \left\{ \begin{pmatrix} d & 0 & 0 \\ 0 & -d & 0 \\ 0 & 0 & 0 \end{pmatrix}, \, \begin{pmatrix} 0 & -d & 0 \\ -d & 0 & 0 \\ 0 & 0 & 0 \end{pmatrix} \right\}.
	\end{aligned}
	\label{eq:raman_tensors}
\end{equation}

In the backscattering geometry along the side surface ($\bar{y}(..)y$), the incident and scattered light propagate along the $y$-axis, confining the polarization vectors to the $x\text{--}z$ ($a\text{--}c$) plane. As the crystal rotates azimuthally by an angle $\theta$ relative to the incident polarization axis, the unit polarization vectors are given by $\hat{\mathbf{e}}_i = (\sin\theta, 0, \cos\theta)^{\mathrm{T}}$, with the analyzer configured as $\hat{\mathbf{e}}_s = \hat{\mathbf{e}}_i$ in the collinear channel ($XX$) and $\hat{\mathbf{e}}_s = (\cos\theta, 0, -\sin\theta)^{\mathrm{T}}$ in the cross-linear channel ($XY$). Evaluating the scattered intensity $I(\theta) \propto \sum_{j} |\hat{\mathbf{e}}_i \cdot \mathcal{R}_j \cdot \hat{\mathbf{e}}_s|^2$ according to Raman selection rules \cite{loudon1964raman,cardona1982light} dictates distinct angular radiation patterns: twofold dumbbell profiles for the $A_{1g}$ and $E_{2g}$ modes alongside a fourfold profile for $E_{1g}$ in the collinear ($XX$) configuration, and fourfold symmetric cloverleaf profiles across all active modes in the cross-linear ($XY$) channel, in full agreement with the calculated polar patterns shown in Fig.~\ref{fig:2}d.

\begin{figure*}[t]
	\centering
	\includegraphics[width=0.8\linewidth]{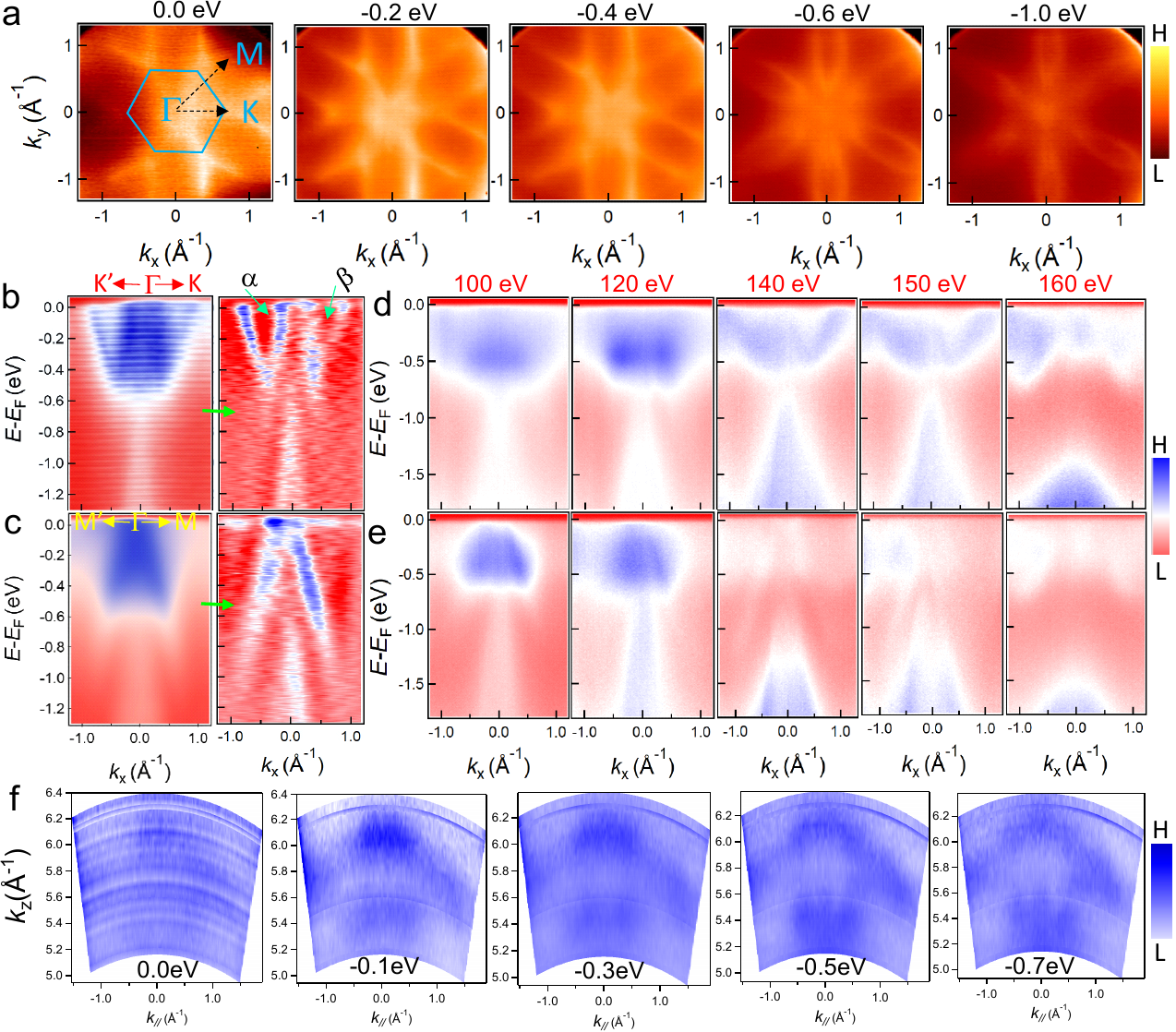}
	\caption{\textbf{Angle-resolved photoemission spectroscopy (ARPES) of $\mathrm{TiAs}$.} 
(a) Constant-energy contour (CEC) maps at $E_{\mathrm{F}}$ and at various binding energies down to $E - E_{\mathrm{F}} = -1.0\text{ eV}$, illustrating the multi-sheet Fermi surface along $K\text{--}\Gamma\text{--}M$; the overlay sketches the two-dimensional hexagonal Brillouin zone. 
(b) Energy--momentum cut along $K\leftarrow\Gamma\rightarrow K$ and the corresponding second-derivative spectrum ($d^2I/dE^2$), resolving the dispersive $\alpha$ and $\beta$ subbands and the saddle-point Van Hove band extremum at $E - E_{\mathrm{F}} \approx -0.5\text{ eV}$ (marked by the green arrow). 
(c) High-symmetry cut along $M\leftarrow\Gamma\rightarrow M$ and its second derivative, displaying steep, monotonic dispersion. The CEC maps and the two high-symmetry cuts were recorded at $h\nu = 130\text{ eV}$. 
(d, e) Incident photon-energy-dependent dispersion cuts along (d) $\Gamma\text{--}K$ and (e) $\Gamma\text{--}M$ across $h\nu = 100\text{--}160\text{ eV}$, tracking the out-of-plane bulk dispersion and spectral weight redistribution. 
(f) Out-of-plane ($k_{\parallel}\text{--}k_z$) constant-energy maps along $K'\text{--}\Gamma\text{--}K$ from $E - E_{\mathrm{F}} = 0.0\text{ eV}$ down to $-0.7\text{ eV}$, evaluated using an inner potential $V_0 \simeq 5\text{ eV}$, demonstrating the transition from quasi-two-dimensional contours near $E_{\mathrm{F}}$ to pronounced $k_z$ dispersion deeper in the valence manifold. All spectra were acquired at $T = 12\text{ K}$ using linear-parallel polarized synchrotron radiation.}
	\label{fig:4}
\end{figure*}

To experimentally validate these theoretical predictions, angle-resolved polarized Raman spectroscopy was conducted on a pristine $\mathrm{TiAs}$ crystal under $532\text{ nm}$ ($2.33\text{ eV}$) excitation in both collinear ($XX$) and cross-linear ($XY$) backscattering geometries at $300\text{ K}$ (Fig.~\ref{fig:2} e--j). The resulting contour maps recorded across azimuthal angles from $0^\circ$ to $360^\circ$ are shown in Figs.~\ref{fig:2}e and \ref{fig:2}f. The measured room-temperature spectrum displays four well-defined first-order optical phonon modes centered at $117$, $224$, $243$, and $280\text{ cm}^{-1}$, in close agreement with the DFPT-calculated frequencies of $116\text{ cm}^{-1}$ ($E_{2g}$), $221\text{ cm}^{-1}$ ($E_{1g}$), $237\text{ cm}^{-1}$ ($E_{2g}$), and $274\text{ cm}^{-1}$ ($A_{1g}$)(Fig.~\ref{fig:2}g). The mean absolute deviation is only 4.2~cm$^{-1}$, with DFPT systematically underestimating the measured frequencies by $1\text{--}3\%$. This small offset is consistent with the lattice expansion and associated phonon softening typical of GGA-PBE calculations~\cite{Perdew1996} and with similar theory--experiment deviations reported for related compounds~\cite{muhammad2026chiralphonon}. Quantitative polar plots extracted for each individual vibrational mode are presented in Figs.~\ref{fig:2}g--j. In the collinear ($XX$) channel, the $E_{2g}(1)$ ($117\text{ cm}^{-1}$), $E_{2g}(2)$ ($243\text{ cm}^{-1}$), and $A_{1g}$ ($280\text{ cm}^{-1}$) modes exhibit distinct twofold dumbbell anisotropies, whereas the $E_{1g}$ ($224\text{ cm}^{-1}$) mode displays fourfold rotational symmetry. In the cross-linear ($XY$) channel, all four optical phonon modes transform into fourfold symmetric cloverleaf patterns. Solid lines in Figs.~\ref{fig:2}c--f represent least-squares fits to the Raman tensor selection rules. Agreement between experimental polar responses and the analytical symmetry predictions corroborates the designated mode assignments and confirms the anisotropic lattice vibrations dictated by the bulk $P6_3/mmc$ symmetry.

While the rotational symmetries of the scattering profiles are strictly governed by the Raman selection rules, the relative peak intensities are strongly modulated by resonance effects. Under $532\text{ nm}$ ($2.33\text{ eV}$) laser excitation, the low-frequency $E_{2g}(1)$ mode at $117\text{ cm}^{-1}$ experiences pronounced resonant enhancement, becoming the predominant feature in the experimental spectrum. First-principles resonant Raman intensity calculations~\cite{hung2024qeraman} yield an intensity distribution that closely mirrors the experimental profile at an excitation energy of $E_L = 2.54\text{ eV}$ (Fig.~\ref{fig:2}c), with the minor energy offset ($\approx 0.21\text{ eV}$) attributable to the well-known DFT bandgap underestimation inherent to semilocal exchange-correlation functionals \cite{perdew1983physical}. In addition to the four fundamental zone-center optical phonons, weak spectral features observed at $153$, $175$, and $206\text{ cm}^{-1}$ are assigned to disorder-activated scattering from non-zone-center states induced by slight off-stoichiometry, whereas faint signals above the one-phonon cutoff frequency ($\sim 303\text{ cm}^{-1}$) originate from second-order overtone and combination phonon processes \cite{cardona1982light}. Together, the first-principles lattice dynamics calculations and polarization-resolved Raman measurements provide a coherent identification of the vibrational landscape and anisotropic lattice response of $\mathrm{TiAs}$.
\begin{figure*}[t]
	\centering
	\includegraphics[width=\linewidth]{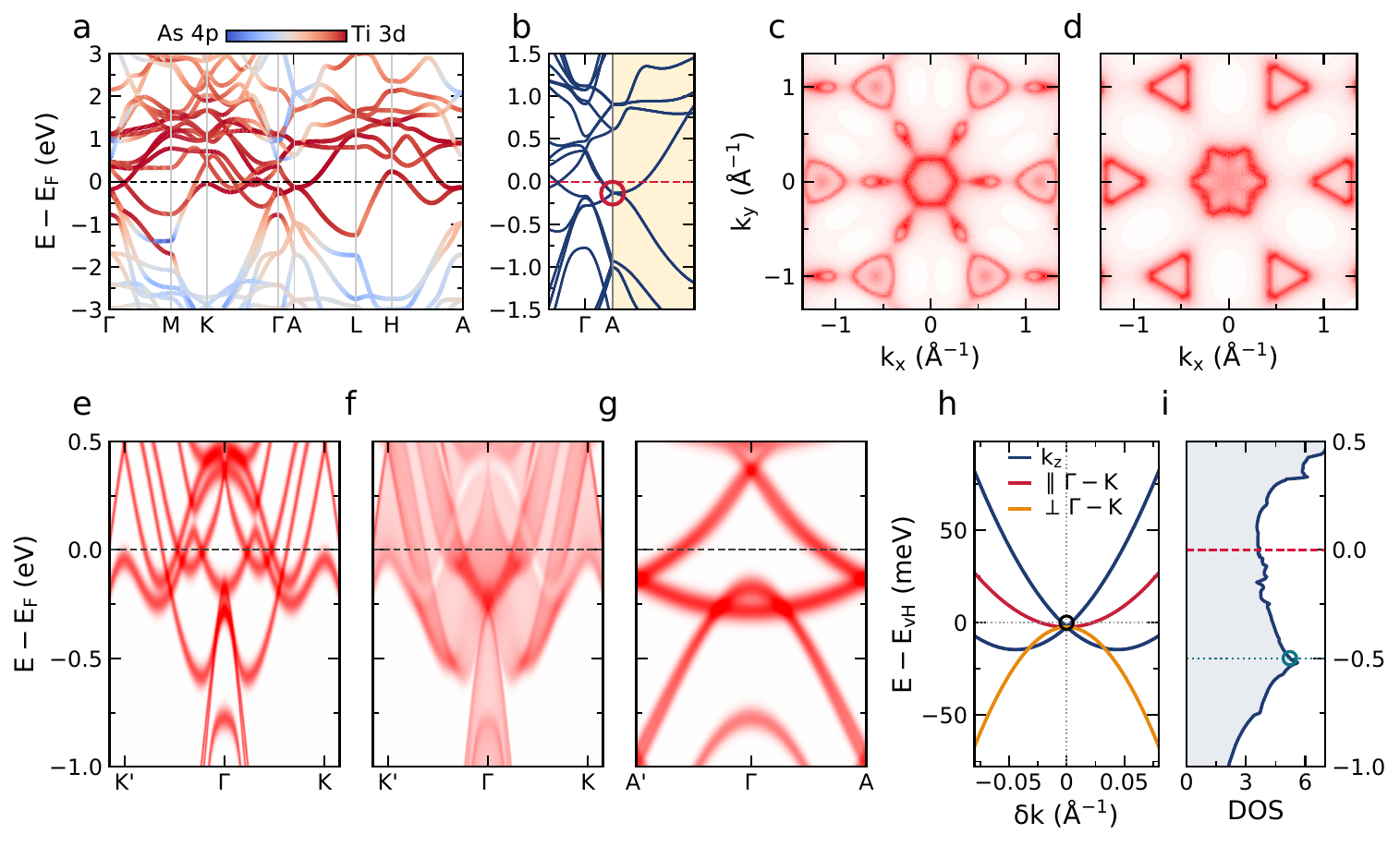}
	\caption{\textbf{Electronic structure, Fermi surface and $k_z$ averaging.}
		(a) Orbital-projected band structure: $\mathrm{Ti}\ 3d$ (red) and $\mathrm{As}\ 4p$ (blue).
		(b) SOC bands near $A$; the shaded $k_z=\pi/c$ boundary hosts screw-symmetry-enforced fourfold degeneracies, including the Dirac-like node at $-0.14$~eV (circle).
		(c,~d) Fermi-surface sections at $k_z=0$ ($h\nu\approx100$~eV) and $\pi/c$ ($h\nu\approx160$~eV), showing the hexagon-to-star evolution.
		(e, f) SOC bulk spectral function along $K'\text{--}\Gamma\text{--}K$ at $k_z = 0$ and averaged over $\Delta k_z \simeq 0.2\ \text{\AA}^{-1}$, illustrating the broadening of sharp crossings by finite $k_z$ resolution.
        (g) Out-of-plane cut $A'\text{--}\Gamma\text{--}A$, where the bands converge in pairs on the zone-boundary planes.
		(h) Dispersions through the van Hove saddle point at $k_z=\pi/c$ ($|k_\parallel|=0.50$~\AA$^{-1}$, $E-E_F=-0.50$~eV; circle), showing opposite in-plane curvatures and screw-protected doublets that split away from the plane.
        (i) Density of states; dashed and dotted lines mark $E_F$ and the saddle-associated maximum at $-0.50$~eV. Panels (b) and (e)–(i) include SOC; (a), (c), and (d) are scalar relativistic.}
	\label{fig:5}
\end{figure*}

Single-crystal $\mathrm{TiAs}$ crystallizes in the nonsymmorphic space group $P6_3/mmc$ ($D_{6h}^4$, No.~194), preserving spatial inversion ($\mathcal{P}$) and time-reversal symmetry ($\mathcal{T}$) alongside a sixfold screw axis $\{C_{6z}|\boldsymbol{\tau}\}$ with fractional translation $\boldsymbol{\tau} = (0, 0, \frac{1}{2})$. In such nonsymmorphic systems, the interplay of $\mathcal{P}\mathcal{T}$ symmetry and fractional translations enforces robust band degeneracies and symmetry-mandated gaplessness across the Brillouin zone boundaries \cite{young2012dirac,bradlyn2017topological}. To uncover how these crystalline symmetries dictate the quasiparticle spectrum and low-energy orbital architecture, high-resolution synchrotron ARPES measurements were conducted at $T = 12\text{ K}$ using linear-parallel polarized radiation and analyzed in direct comparison with relativistic first-principles calculations incorporating spin-orbit coupling (SOC) [Figs.~\ref{fig:4} and \ref{fig:5}].

Figure~\ref{fig:4}(a) displays constant-energy contour (CEC) maps collected at $h\nu = 130\text{ eV}$ across binding energies from the chemical potential ($E - E_{\mathrm{F}} = 0.0\text{ eV}$) down to $-1.0\text{ eV}$. At the Fermi level, the spectral weight delineates a multi-sheet Fermi surface conforming to the sixfold rotational ($C_6$) symmetry of the hexagonal basal plane, consisting of a central pocket enclosing $\bar{\Gamma}$ surrounded by outward-dispersing sheets oriented along the high-symmetry $\bar{\Gamma}\text{--}\bar{K}$ and $\bar{\Gamma}\text{--}\bar{M}$ azimuths. With increasing binding energy, the constant-energy contours undergo a topological reconfiguration: the central pocket expands while the outer lobes pinch off and detach, transitioning from an interconnected hexagonal geometry at $E_{\mathrm{F}}$ into isolated, star-like petals at deeper binding energies. This reconfiguration signifies a Lifshitz transition in contour connectivity, where the pinching off and detachment of adjacent sheets is driven by directional in-plane hopping within the $\mathrm{Ti}\ 3d$ manifold.

\begin{figure*}[t]
	\centering
	\includegraphics[width=0.8\linewidth]{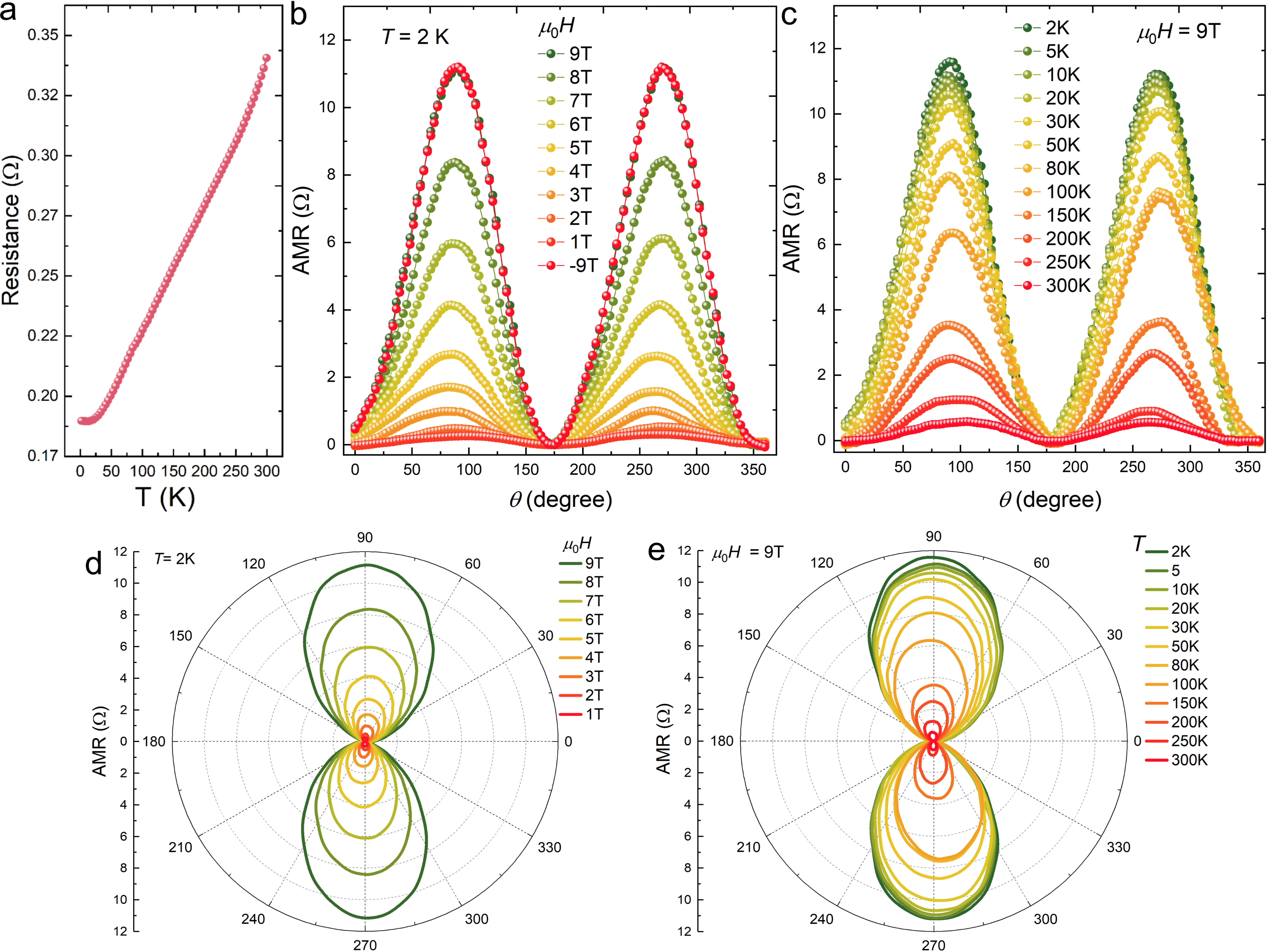}
	\caption{\textbf{Temperature- and angle-dependent electrical transport.} 
		(a) Longitudinal resistance from $300$ to $2\text{ K}$, showing monotonic metallic conduction. 
		(b) Angular magnetoresistance (AMR) versus field-rotation angle $\theta$ at $T = 2\text{ K}$ for fields from $\mu_0 H = 1$ to $9\text{ T}$ (and $-9\text{ T}$). The response is twofold ($C_2$), with minima at $\theta = 0^\circ, 180^\circ$ ($\mathbf{H} \parallel \mathbf{I}$) and maxima at $90^\circ, 270^\circ$ ($\mathbf{H} \perp \mathbf{I}$). 
		(c) The same at fixed $\mu_0 H = 9\text{ T}$ from $2$ to $300\text{ K}$. 
		(d,~e) Polar representations of the AMR, showing the dumbbell-shaped twofold profile versus (d) field at $2\text{ K}$ and (e) temperature at $9\text{ T}$.}
	\label{fig:6}
\end{figure*}

To resolve the detailed band dispersions and illuminate the in-plane electronic anisotropy, energy--momentum cuts and second-derivative spectra ($d^2I/dE^2$) along $K\leftarrow\Gamma\rightarrow K$ and $M\leftarrow\Gamma\rightarrow M$ are compared in Figs.~\ref{fig:4}(b) and \ref{fig:4}(c). Along the $K\leftarrow\Gamma\rightarrow K$ trajectory [Fig.~\ref{fig:4}(b)], two dispersive subbands, denoted $\alpha$ and $\beta$, cross the chemical potential with steep Fermi velocities. Deeper in the valence manifold, the second-derivative spectrum unambiguously resolves a local band extremum at $E - E_{\mathrm{F}} \approx -0.5\text{ eV}$ [green horizontal arrow in Fig.~\ref{fig:4}(b)], where the band flattens horizontally with a vanishing in-plane group velocity ($\nabla_{\mathbf{k}_{\parallel}} E \approx 0$). In stark contrast, along the orthogonal $M\leftarrow\Gamma\rightarrow M$ path [Fig.~\ref{fig:4}(c)], the corresponding band branches disperse steeply and monotonically through this identical binding-energy window [green arrow in Fig.~\ref{fig:4}(c)] without forming an energy turnaround or dispersion plateau. This marked momentum contrast—characterized by a vanishing in-plane group velocity along $\Gamma\text{--}K$ versus steep dispersion along $\Gamma\text{--}M$—provides clear evidence of strong in-plane anisotropy and, in conjunction with the neck-pinching contour topology in Fig.~\ref{fig:4}(a), identifies an in-plane Van Hove saddle point. Physically, this local saddle point acts as the momentum-space locus mediating the neck-collapsing Lifshitz transition~\cite{Lifshitz1960}, driven by directional, orbital-selective hybridization between $\mathrm{Ti}\ 3d$ and $\mathrm{As}\ 4p$ states. Crucially, as evidenced by the photon-energy-dependent dispersion series in Fig.~\ref{fig:4}(d), this valence-band extremum persists robustly across the entire excitation window ($h\nu = 100\text{--}160\text{ eV}$) rather than manifesting as an isolated resonance. This systematic persistence confirms that the saddle-like topology is an intrinsic property of the bulk electronic manifold rather than a photoemission matrix-element artifact or trivial surface resonance~\cite{Sobota2021RMP, Damascelli2003RMP}.

The three-dimensional nature of the electronic states was investigated by tracking the out-of-plane dispersion across incident photon energies from $h\nu = 100\text{ eV}$ to $160\text{ eV}$ along both high-symmetry cuts [Figs.~\ref{fig:4}(d) and \ref{fig:4}(e)]. Along both trajectories, the pocket dimensions, Fermi wavevectors, and band velocities exhibit systematic modulations with excitation energy, confirming that the photoemission signal originates from bulk electronic states rather than trivial two-dimensional surface terminations. In particular, tuning from $h\nu \approx 100\text{ eV}$ toward $160\text{ eV}$ induces a marked redistribution of spectral weight, accompanied by an energy downshift and an avoided-crossing hybridization gap near deeper binding energies.

Using the standard free-electron final-state model, $k_z = \sqrt{2m_0(E_{\mathrm{kin}}\cos^2\theta + V_0)}/\hbar$, with an experimentally determined inner potential $V_0 \simeq 5\text{ eV}$, out-of-plane momentum tomographies ($k_{\parallel}\text{--}k_z$) were reconstructed across successive Brillouin zones [Fig.~\ref{fig:4}(f)]. At the Fermi level ($E - E_{\mathrm{F}} = 0.0\text{ eV}$), the $k_{\parallel}\text{--}k_z$ contours exhibit nearly straight, cylindrical sheets along $k_z$, reflecting a quasi-two-dimensional character of the Fermi surface sheets near the chemical potential. In contrast, deeper valence states ($E - E_{\mathrm{F}} \le -0.3\text{ eV}$) develop strong, periodic modulations along $k_z$ that faithfully match the bulk Brillouin zone periodicity. This continuous evolution from quasi-cylindrical Fermi sheets to strongly undulating bulk valence manifolds confirms the three-dimensional nature of the bulk electronic structure.

\begin{figure*}[t]
	\centering
	\includegraphics[width=0.8\linewidth]{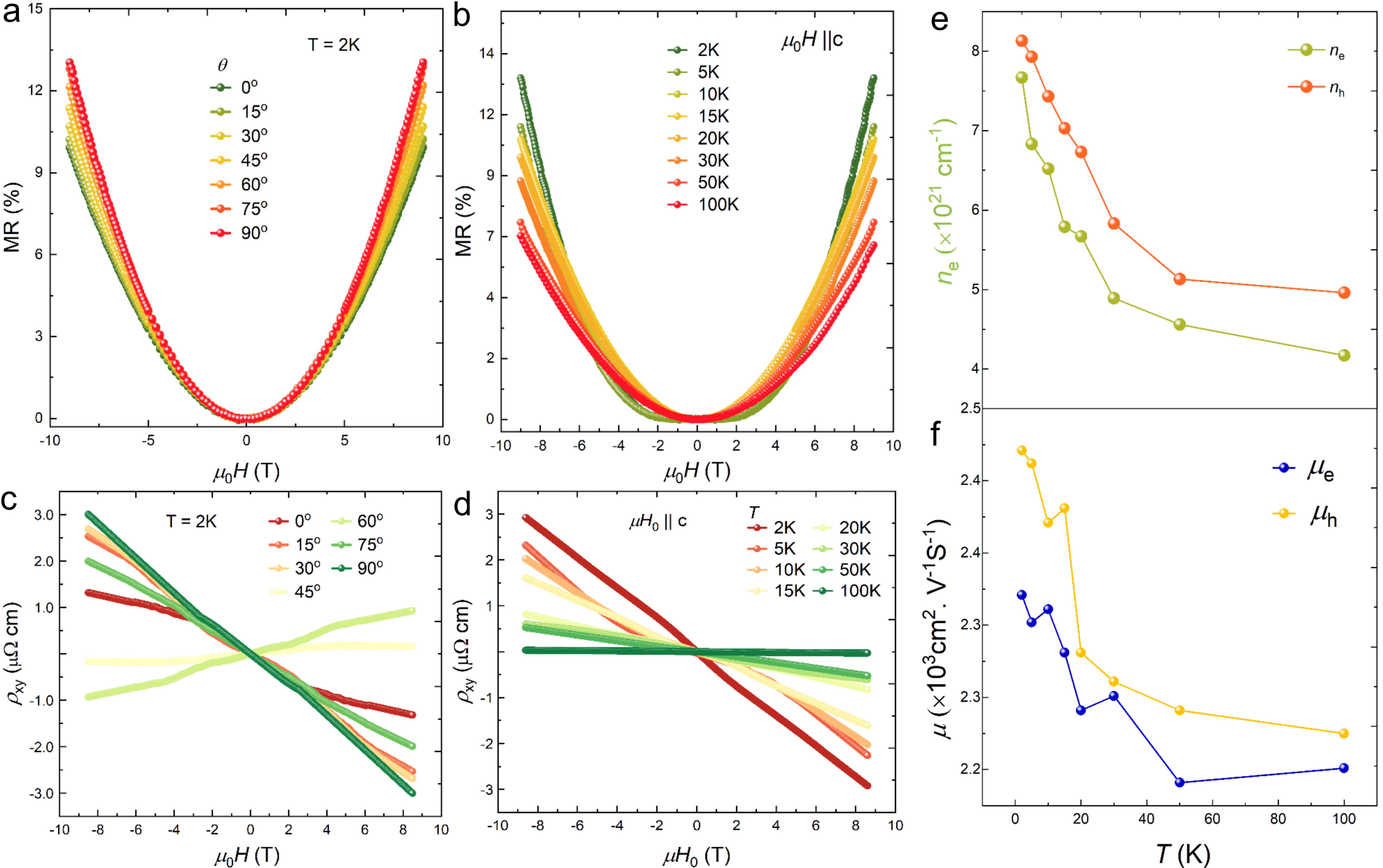}
	\caption{\textbf{Magnetotransport, Hall effect and two-carrier analysis.} 
		(a) Transverse magnetoresistance, $\mathrm{MR} = [\rho_{xx}(H) - \rho_{xx}(0)]/\rho_{xx}(0)$, up to $9\text{ T}$ at $T = 2\text{ K}$ for $\theta = 0^\circ$ ($\mathbf{H} \parallel \mathbf{I}$) to $90^\circ$ ($\mathbf{H} \perp \mathbf{I}$). 
		(b) The same with $\mu_0\mathbf{H} \parallel \mathbf{c}$ from $2$ to $100\text{ K}$; the non-saturating, quadratic-like MR reaches $\sim 13\%$ at $2\text{ K}$. 
		(c) Hall resistivity $\rho_{xy}$ at $T = 2\text{ K}$ for the same angles $\theta$, the low-field slope changing sign from negative to positive. 
		(d) Nonlinear $\rho_{xy}(H)$ with $\mu_0\mathbf{H} \parallel \mathbf{c}$ from $2$ to $100\text{ K}$. 
		(e,~f) Temperature dependence of (e) the electron ($n_e$) and hole ($n_h$) densities and (f) their mobilities ($\mu_e, \mu_h$), from a simultaneous fit to the semiclassical two-carrier model.}
	\label{fig:7}
\end{figure*}

Quantitative validation of these spectroscopic signatures is provided by relativistic density functional theory (DFT) calculations incorporating spin-orbit coupling [Fig.~\ref{fig:5}]. The orbital-projected band structure [Fig.~\ref{fig:5}(a)] shows that the low-energy manifold within $\pm 1\text{ eV}$ of $E_{\mathrm{F}}$ is dominated by $\mathrm{Ti}\ 3d$ states, whereas filled $\mathrm{As}\ 4p$ states reside primarily below $-1.5\text{ eV}$. Near the zone center ($\Gamma$), an inverted $\mathrm{As}\ 4p$ band disperses upward into the $\mathrm{Ti}\ 3d$ manifold, producing an orbital band inversion that underpins the multi-band crossings resolved experimentally near $-0.5\text{ eV}$ in Fig.~\ref{fig:4}(b). The coexistence of inversion ($\mathcal{P}$) and time-reversal ($\mathcal{T}$) symmetries allows the global band topology of $\mathrm{TiAs}$ to be classified by parity eigenvalues of occupied spinor states across the eight time-reversal-invariant momenta (TRIM)~\cite{fu2007topological}. For the 108 occupied valence bands, the even/odd parity counts evaluate to $n^+/n^- = 68/40$ at $\Gamma$ and $62/46$ at $M$. At the zone-boundary TRIM points ($A$ and the three $L$ points), the nonsymmorphic $6_3$ screw symmetry dictates fourfold-degenerate multiplets possessing vanishing net inversion character, contributing neutrally to the parity sum. Evaluating the symmetry indicator $\mathbb{Z}_4 = \sum_{\mathrm{TRIM}} (n^+ - n^-)/2 \pmod 4$ yields $\mathbb{Z}_4 = 2$~\cite{po2017symmetry}, corresponding to the strong topological index $\nu_0 = 1$ and precluding an atomic-insulator description. Furthermore, band compatibility relations along the $\Gamma\text{--}A$ line cannot be satisfied by an isolated set of filled bands, rigorously confirming $\mathrm{TiAs}$ as a symmetry-enforced topological semimetal~\cite{bradlyn2017topological,vergniory2019complete}.

In addition to the topological indicators, the nonsymmorphic $\{C_{6z}|\boldsymbol{\tau}\}$ screw symmetry enforces pairwise band sticking across the entire $k_z = \pi/c$ ($A\text{--}L\text{--}H$) zone-boundary plane, yielding fourfold-degenerate Dirac nodes [Fig.~\ref{fig:5}(b)]. The lowest-energy crossing is pinned at the $A$ point at $E - E_{\mathrm{F}} = -0.14\text{ eV}$ [red circle in Fig.~\ref{fig:5}(b)]. This symmetry-enforced band evolution provides a direct microscopic basis for our photon-energy-dependent ARPES data: calibrated with $V_0 \simeq 5\text{ eV}$, $h\nu \approx 100\text{ eV}$ selectively probes the vicinity of the $k_z = 0$ plane, while $h\nu \approx 160\text{ eV}$ samples near the $k_z = \pi/c$ boundary. The calculated valence-band apex shifts from $-0.28\text{ eV}$ at $\Gamma$ down to $-0.97\text{ eV}$ at $A$, accurately reproducing the experimentally resolved energy downshift and avoided-crossing gap between $100\text{ eV}$ and $160\text{ eV}$ [Figs.~\ref{fig:4}(d) and \ref{fig:4}(e)]. Concurrently, the calculated constant-energy Fermi surface sections reproduce the observed topological crossover from an interconnected hexagon at $k_z = 0$ to isolated, star-like petals at $k_z = \pi/c$ [Figs.~\ref{fig:5}(c) and \ref{fig:5}(d), in excellent agreement with the experimental CEC evolution [Fig.~\ref{fig:4}(a)].

The pronounced out-of-plane dispersion also accounts for the intrinsic linewidths and spectral weight distribution in ARPES. Along $\Gamma\text{--}K$, bands crossing $E_{\mathrm{F}}$ exhibit an out-of-plane dispersion shift of $0.14\text{--}0.60\text{ eV}$ between $k_z = 0$ and $k_z = \pi/c$. Along the out-of-plane cut $A'\text{--}\Gamma\text{--}A$ [Fig.~\ref{fig:5}(g)], the bands disperse strongly and converge pairwise at the boundary planes. Due to the finite photoelectron mean free path, the experimental out-of-plane momentum resolution is $\Delta k_z \simeq 0.2\text{ \AA}^{-1}$. Consequently, ARPES spectra inherently integrate spectral weight over an extended $k_z$ window. While calculations evaluated at a single $k_z = 0$ cut exhibit sharp, discrete crossings [Fig.~\ref{fig:5}(e)], the $k_z$-broadened bulk spectral function [Fig.~\ref{fig:5}(f)] displays diffuse, widened bands that faithfully reproduce the experimental intensity distribution.

Finally, our calculations unambiguously corroborate the three-dimensional Van Hove saddle points in the band structure. The first resides near the $K$ point, situated $47\text{ meV}$ below the chemical potential. The second, more prominent saddle point lies on the $k_z = \pi/c$ plane at $E - E_{\mathrm{F}} \simeq -0.50\text{ eV}$ with $|\mathbf{k}_{\parallel}| = 0.50\text{ \AA}^{-1}$ [Fig.~\ref{fig:5}(h)], where the $\{C_{6z}|\boldsymbol{\tau}\}$ screw symmetry enforces a fourfold degeneracy. Orthogonal momentum cuts centered on this critical point demonstrate opposite signs of band curvature ($\partial^2 E / \partial k_i^2$) along the in-plane trajectories ($\parallel \Gamma\text{--}K$ vs $\perp \Gamma\text{--}K$) and along the out-of-plane $k_z$ path [Fig.~\ref{fig:5}(h)], establishing a genuine 3D saddle-point singularity. Because 3D saddle points produce continuous slope discontinuities rather than divergent logarithmic spikes, the calculated total DOS remains continuous near $E_{\mathrm{F}}$ with a finite value of $N(E_{\mathrm{F}}) = 3.7\text{ states/eV per unit cell}$ [Fig.~\ref{fig:5}(i)], explaining why the shallow state at $-47\text{ meV}$ does not trigger structural or electronic instabilities. In contrast, the zone-boundary saddle point at $-0.50\text{ eV}$ coincides with a prominent local peak in the occupied DOS [Fig.~\ref{fig:5}(i)]. This directly confirms that the vanishing group velocity ($\nabla_{\mathbf{k}_{\parallel}} E \approx 0$) and band turnaround resolved in ARPES second derivatives [Fig.~\ref{fig:4}(b)] originate from this symmetry-protected 3D Van Hove saddle point.

To establish the macroscopic manifestations of the multiband Fermi surface and relativistic carrier dynamics deduced from ARPES and DFT, temperature- and field-dependent electrical transport measurements were performed on $\mathrm{TiAs}$ single crystals [Figs.~\ref{fig:6} and \ref{fig:7}]. As shown in Fig.~\ref{fig:6}(a), the zero-field longitudinal resistance decreases monotonically upon cooling from $300\text{ K}$ down to $2\text{ K}$ without anomalous kinks or hysteresis, demonstrating robust metallic conduction free from structural or magnetic phase transitions. The high residual resistivity ratio $[\mathrm{RRR} = R(300\text{ K})/R(2\text{ K}) \approx 2.0]$ reflects clean metallic conduction where low-temperature transport is governed primarily by residual defect scattering rather than thermal phonon dissipation.

To probe the directional anisotropy of the scattering channels and the out-of-plane Fermi-surface topology, angular magnetoresistance (AMR) was systematically recorded by rotating the magnetic field within the plane containing the current axis and the crystallographic $c$ axis [Fig.~\ref{fig:6}(b)]. At $T = 2\text{ K}$, the AMR exhibits a pronounced twofold ($C_2$) symmetric oscillation. The resistance reaches its global minimum at $\theta = 0^\circ$ and $180^\circ$ where the magnetic field aligns parallel to the current path ($\mathbf{H} \parallel \mathbf{I}$, longitudinal geometry), minimizing orbital Lorentz deflection. Conversely, rotating the field toward $\theta = 90^\circ$ and $270^\circ$ establishes the transverse geometry ($\mathbf{H} \perp \mathbf{I}$), which maximizes the Lorentz force $\mathbf{F} = -e(\mathbf{v} \times \mathbf{B})$ and drives carrier trajectories across the anisotropic cyclotron orbits of the Fermi surface, yielding a prominent resistance maximum ($\sim 11.5\,\Omega$ at $9\text{ T}$).

The temperature evolution of the AMR at $\mu_0 H = 9\text{ T}$ [Fig.~\ref{fig:6}(c)] demonstrates that while the twofold symmetry persists up to $300\text{ K}$, the oscillation amplitude attenuates systematically with thermal activation. In polar representations [Figs.~\ref{fig:6}(d) and \ref{fig:6}(e)], this response forms sharp, dumbbell-shaped lobes oriented along the transverse axis ($\theta = 90^\circ/270^\circ$). This prominent anisotropy is a direct transport consequence of the corrugated, open-orbit topology along $k_z$ identified in our photon-energy-dependent ARPES [Fig.~\ref{fig:4}(f)] and DFT band dispersions [Fig.~\ref{fig:5}(g)]. Above $150\text{ K}$, the rapid contraction of the dumbbell lobes signals that inelastic thermal phonon scattering overwhelms coherent cyclotron orbits.

To further disentangle the individual conduction channels, field-dependent magnetoresistance $[\mathrm{MR} = (\rho_{xx}(H) - \rho_{xx}(0))/\rho_{xx}(0)]$ and Hall resistivity ($\rho_{xy}$) were investigated [Fig.~\ref{fig:7}]. As illustrated in Figs.~\ref{fig:7}(a) and \ref{fig:7}(b), the transverse MR exhibits a positive, non-saturating quadratic-like field dependence up to $9\text{ T}$, reaching $\sim 13\%$ at $2\text{ K}$ under $\theta = 90^\circ$ ($\mathbf{H} \perp \mathbf{I}$). The MR amplitude scales monotonically with the tilting angle toward the transverse configuration [Fig.~\ref{fig:7}(a)] and decreases steadily with warming [Fig.~\ref{fig:7}(b)], consistent with the thermal suppression of carrier mobilities.

The field-dependent Hall resistivity profiles ($\rho_{xy}$) provide direct evidence for multiband conduction. At $T = 2\text{ K}$ across varying field orientations [Fig.~\ref{fig:7}(c)], $\rho_{xy}$ exhibits pronounced nonlinear field dependence. Strikingly, tilting the magnetic field from $\theta = 0^\circ$ ($\mathbf{H} \parallel \mathbf{I}$) to $\theta = 90^\circ$ ($\mathbf{H} \parallel \mathbf{c} \perp \mathbf{I}$) induces a sign reversal in the low-field slope ($d\rho_{xy}/dH$) from negative to positive. This orientation-dependent slope inversion demonstrates that while high-mobility electron carriers dominate low-field transport when field components lie within the basal plane, hole pockets govern the out-of-plane transport along the $c$ axis. At fixed orientation ($\mu_0\mathbf{H} \parallel \mathbf{c}$), the nonlinear Hall curvature persists across temperatures [Fig.~\ref{fig:7}(d)], confirming the continuous participation of coexisting electron and hole pockets up to $100\text{ K}$.

To extract quantitative parameters for these coexisting carrier channels, the field dependence of $\rho_{xy}$ and the longitudinal conductivity were simultaneously fitted to the semiclassical two-carrier model~\cite{ali2014large}:
\begin{equation}
	\rho_{xy} = \frac{\mu_0 H}{e} \frac{(n_h \mu_h^2 - n_e \mu_e^2) + (n_h - n_e) \mu_e^2 \mu_h^2 (\mu_0 H)^2}{(n_e \mu_e + n_h \mu_h)^2 + (n_h - n_e)^2 \mu_e^2 \mu_h^2 (\mu_0 H)^2},
	\label{eq:two_carrier}
\end{equation}
where $n_e$ ($n_h$) and $\mu_e$ ($\mu_h$) denote the carrier concentrations and mobilities of electrons and holes, respectively. The temperature-dependent evolution of these parameters is summarized in Figs.~\ref{fig:7}(e) and \ref{fig:7}(f). At $2\text{ K}$, the extracted carrier densities are on the order of $\sim 10^{21}\text{ cm}^{-3}$, consistent with high-density metallic conduction. Specifically, the hole concentration ($n_h \approx 8.1 \times 10^{21}\text{ cm}^{-3}$) slightly exceeds the electron concentration ($n_e \approx 7.6 \times 10^{21}\text{ cm}^{-3}$), proving that $\mathrm{TiAs}$ is an uncompensated multiband metal ($n_h \ne n_e$). This carrier imbalance naturally accounts for the moderate magnetoresistance ($\sim 13\%$), in sharp contrast to the extreme, non-saturating giant magnetoresistance observed in perfectly compensated semimetals such as $\mathrm{WTe_2}$ and $\mathrm{ZrAs_2}$~\cite{ali2014large,wang2016origin,hossain2025superconductivity,muhammad2024electronic,huang2015observation,naumann2020orbital}. At $2\text{ K}$, both carrier mobilities are comparable, with $\mu_h \approx 2.5 \times 10^3\text{ cm}^2\text{V}^{-1}\text{s}^{-1}$ and $\mu_e \approx 2.2 \times 10^3\text{ cm}^2\text{V}^{-1}\text{s}^{-1}$ [Fig.~\ref{fig:7}(f)], decaying smoothly with increasing temperature due to acoustic and optical phonon scattering. These transport findings provide macroscopic corroboration that the low-energy physics of $\mathrm{TiAs}$ is governed by an uncompensated, multiband Fermi surface populated by relativistic, orbital-hybridized $\mathrm{Ti}\ 3d$ carriers.

\section{Conclusion}

In summary, we establish stoichiometric $\mathrm{TiAs}$ as a model symmetry-enforced topological semimetal within the $3d$ transition-metal monopnictide family. Polarized Raman spectroscopy and phonon calculations confirm the dynamical stability and complete vibrational symmetry of the hexagonal $P6_3/mmc$ lattice. Synchrotron ARPES and relativistic DFT calculations uncover an orbital-dependent dimensionality crossover, evolving from quasi-two-dimensional Fermi cylinders into a strongly corrugated three-dimensional valence manifold along $k_z$. Analysis of symmetry indicators confirms a nontrivial classification ($\mathbb{Z}_4=2$, $\nu_0=1$), hosting nonsymmorphic screw-symmetry-enforced fourfold Dirac degeneracies pinned at the $k_z = \pi/c$ plane. Furthermore, energy-dependent contours resolve an in-plane, neck-collapsing Lifshitz transition mediated by a Van Hove saddle point at $E - E_{\mathrm{F}} \approx -0.5\text{ eV}$ along $\Gamma\text{--}K$, characterized by vanishing group velocity, opposite orthogonal curvatures, and an occupied density-of-states peak. In magnetotransport, open-orbit cyclotron trajectories along the corrugated $k_z$ sheets produce a twofold ($C_2$) dumbbell-shaped angular magnetoresistance, while two-carrier modeling quantifies uncompensated hole-dominated transport ($n_h > n_e$) that explains both the moderate magnetoresistance and the field-orientation-induced Hall slope inversion. These results establish $\mathrm{TiAs}$ as a benchmark platform linking nonsymmorphic Dirac topology, Van Hove saddle points, and anisotropic electronic transport.

\section{Experiments}

\subsection{X-Ray Diffraction and Structural Refinement}
Room-temperature powder x-ray diffraction (XRD) patterns were collected on pulverized single crystals using a commercial diffractometer with $\mathrm{Cu}\ K\alpha$ radiation ($\lambda = 1.5406\text{ \AA}$). The structural parameters, phase purity, and atomic occupancies were refined via the Rietveld method utilizing the MAUD (Material Analysis Using Diffraction) software package. An isotropic crystallite size model and background polynomials were refined alongside unit-cell dimensions and atomic displacement parameters ($B_{\mathrm{iso}}$), yielding low reliability factors ($R_{\mathrm{wp}}$, $R_{\mathrm{p}}$, and $\chi^2$) that confirm single-phase hexagonal $\mathrm{TiAs}$ without detectable secondary binaries.

\subsection{Raman Scattering Spectroscopy}
Angle-resolved polarized Raman spectra were recorded at room temperature in a backscattering geometry using continuous-wave $532\text{ nm}$ laser excitation. A half-wave ($\lambda/2$) plate positioned directly above the objective lens rotated the incident and scattered polarization vectors simultaneously relative to the sample crystallographic axes. Spectra were collected in both co-polarized [$XX$, $\bar{y}(zz)y$] and cross-polarized [$XY$, $\bar{y}(zx)y$] configurations, where the analyzer was set parallel and perpendicular to the incident beam polarization, respectively. The propagation direction of both incident and scattered light was defined along the crystallographic $y$-axis, with polarization geometries tracking the orthogonal $x$- and $z$-directions.

\subsection{Magnetotransport Measurements}
Electrical transport properties were evaluated using a 16~T Physical Property Measurement System (PPMS, Quantum Design) at the High Magnetic Field Laboratory (HMFL), Chinese Academy of Sciences (CAS), Hefei, China. Ohmic contacts were prepared in a standard four-probe configuration on polished single crystals using gold wires affixed with conductive silver epoxy. Angular magnetoresistance (AMR), longitudinal resistivity ($\rho_{xx}$), and Hall resistivity ($\rho_{xy}$) were systematically recorded from $2\text{ K}$ to $300\text{ K}$ under external magnetic fields up to $9\text{ T}$. Field-orientation-dependent measurements were enabled using a motorized horizontal rotator puck to systematically adjust the angle $\theta$ between the applied field and the sample orientation.

\subsection{Angle-Resolved Photoemission Spectroscopy (ARPES)}
Synchrotron ARPES measurements were conducted at the high-resolution BL03U beamline of the Shanghai Synchrotron Radiation Facility (SSRF), equipped with a Scienta DA30L hemispherical electron energy analyzer. High-quality single crystals were cleaved \textit{in situ} along the $(001)$ plane under ultrahigh-vacuum conditions with a base pressure below $5 \times 10^{-11}\text{ mbar}$. Data were acquired at a base temperature of $12\text{ K}$ over a photon energy range of $90\text{--}160\text{ eV}$ using horizontally polarized synchrotron radiation. The overall energy and angular resolutions were set to better than $13\text{ meV}$ and $0.1^\circ$, respectively.

\subsection{Computational Methods}
First-principles electronic structure calculations were performed within the framework of density functional theory (DFT) using the \textsc{Quantum ESPRESSO} package~\cite{Giannozzi2017}. The generalized gradient approximation (GGA) parameterized by Perdew, Burke, and Ernzerhof (PBE)~\cite{Perdew1996} was employed alongside norm-conserving Optimized Norm-Conserving Vanderbilt (ONCV) pseudopotentials~\cite{hamman2013}. The kinetic energy cutoff for the plane-wave basis was set to $80\text{ Ry}$, with a Methfessel--Paxton smearing width of $0.01\text{ Ry}$~\cite{Methfessel1989}. The relaxed hexagonal $P6_3/mmc$ unit cell parameters ($a = 3.646\text{ \AA}$, $c = 12.101\text{ \AA}$) were adopted throughout the calculations. 
Relativistic calculations including spin--orbit coupling (SOC) used fully relativistic ONCV pseudopotentials. We obtained space-group irreducible representations at maximal $\mathbf{k}$ points using \texttt{irvsp}/\texttt{IR2PW}~\cite{gao2021irvsp} and \texttt{irrep}~\cite{iraola2022irrep}, and evaluated the symmetry indicator $\mathbb{Z}_4$ from the inversion eigenvalues of the 108 occupied spinor bands~\cite{po2017symmetry}. We determined the band connectivity independently from the corresponding compatibility relations~\cite{bradlyn2017topological}. Bulk spectral functions were calculated from a maximally localized Wannier Hamiltonian constructed with \textsc{Wannier90}~\cite{Pizzi2020} using Ti $3d$ and As $4p$ projections on a $12\times12\times4$ $\mathbf{k}$ mesh; the $k_z$-averaged spectrum was broadened by $1/\lambda$ along $k_z$. Band critical points were identified from the energy gradients and curvatures, and the SOC density of states was evaluated with the optimized tetrahedron method~\cite{kawamura2014tetrahedron} on a $64\times64\times24$ $\mathbf{k}$ mesh. Phonons and vibrational eigenmodes were calculated within density-functional perturbation theory (DFPT)~\cite{Baroni2001} on a $6\times6\times3$ $\mathbf{q}$ mesh, and resonant Raman intensities with \textsc{QERaman}~\cite{hung2024qeraman}.

\section{ACKNOWLEDGMENTS}
The project was supported by National Natural Science Foundation of China (Nos. W2632012, 1261101385, T2394475, and 52261145694). Part of the Research at the National Synchrotron Radiation Centre SOLARIS is supported by the Ministry of Science and Higher Education, Poland, under contract no. 1/SOL/2021/2. Part of this research used Shanghai Synchrotron Radiation Facility (SSRF) of BL03U(31124.02.SSRF.BL03U) for the assistance on ARPES measurements. We thank the Steady High Magnetic Field Facility (https://cstr.cn/31125.02.SHMFF.XRD) of CAS for data collection and analysis. S.M. acknowledges the Texas Advanced Computing Center (TACC) at The University of Texas at Austin (http://www.tacc.utexas.edu) for providing computational resources through the ACCESS program under award No.~TG-PHY250052.

\bibliography{biblio}

@article{maud_ref,
	title = {Combined structure and texture analysis of thin films by energy-dispersive and combined diffractometry using {MAUD}},
	author = {Lutterotti, Luca and Matthies, Siegfried and Wenk, Hans-Rudolf},
	journal = {Proc. 12th Int. Conf. Textures of Materials (ICOTOM-12)},
	volume = {1},
	pages = {1599},
	year = {1999}
}

@article{march_dollase,
title={Correction of intensities for preferred orientation in powder diffractometry: application of the March model},
author={Dollase, WA},
journal = {J. Appl. Crystallogr.},
volume={19},
number={4},
pages={267--272},
year={1986},
publisher={International Union of Crystallography},
	doi     = {10.1107/S0021889886089458},
	url     = {https://doi.org/10.1107/S0021889886089458}
}

@article{bachmayer1955,
	title={Die Struktur von TiAs},
	author={Bachmayer, K and Nowotny, Helmut and Kohl, A},
	journal = {Monatsh. Chem.},
	volume={86},
	number={1},
	pages={39--43},
	year={1955},
	publisher={Springer},
	doi     = {10.1007/BF00899271},
	url     = {https://doi.org/10.1007/BF00899271}
}

@article{mp1822,

title={Commentary: The Materials Project: A materials genome approach to accelerating materials innovation},
author={Jain, Anubhav and Ong, Shyue Ping and Hautier, Geoffroy and Chen, Wei and Richards, William Davidson and Dacek, Stephen and Cholia, Shreyas and Gunter, Dan and Skinner, David and Ceder, Gerbrand and others},
journal = {APL Mater.},
volume={1},
number={1},
pages={011002},
year={2013},
publisher={AIP Publishing},
	doi     = {10.1063/1.4812323},
	url     = {https://doi.org/10.1063/1.4812323}
}

@article{kjekshus1964,
	author  = {Kjekshus, Arne and Pearson, William B.},
	title   = {Phases with the nickel arsenide and closely related structures},
	journal = {Prog. Solid State Chem.},
	volume  = {1},
	pages   = {83--174},
	year    = {1964},
	doi     = {10.1016/0079-6786(64)90005-7},
	url     = {https://doi.org/10.1016/0079-6786(64)90005-7}
}

@book{moulder1992xps,
	title     = {Handbook of X-ray Photoelectron Spectroscopy},
	author    = {Moulder, John F. and Stickle, William F. and Sobol, Peter E. and Bomben, Kenneth D.},
	editor    = {Chastain, Jill},
	publisher = {Perkin-Elmer Corporation},
	address   = {Eden Prairie, Minnesota},
	year      = {1992}
}

@article{Baroni2001,
  title = {{Phonons and related crystal properties from density-functional perturbation theory}},
  author = {Baroni, Stefano and de Gironcoli, Stefano and Dal Corso, Andrea and Giannozzi, Paolo},
  journal = {Rev. Mod. Phys.},
  volume = {73},
  issue = {2},
  pages = {515--562},
  numpages = {0},
  year = {2001},
  month = {Jul},
  publisher = {American Physical Society},
  doi = {10.1103/RevModPhys.73.515},
  url     = {https://doi.org/10.1103/RevModPhys.73.515}
}

@article{fateley1971infrared,
	title   = {Infrared and Raman Selection Rules for Lattice Vibrations: The Correlation Method},
	author  = {Fateley, W. G. and McDevitt, Neil T. and Bentley, Freeman F.},
	journal = {Appl. Spectrosc.},
	volume  = {25},
	number  = {2},
	pages   = {155--173},
	year    = {1971},
	doi     = {10.1366/000370271779948600},
	url     = {https://doi.org/10.1366/000370271779948600}
}

@article{kroumova2003bilbao,
	title   = {Bilbao Crystallographic Server: useful databases and tools for phase-transition studies},
	author  = {Kroumova, E. and Aroyo, M. I. and Perez-Mato, J. M. and Kirov, A. and Capillas, C. and Ivantchev, S. and Wondratschek, H.},
	journal = {Phase Transit.},
	volume  = {76},
	number  = {1-2},
	pages   = {155--170},
	year    = {2003},
	doi     = {10.1080/0141159031000076110},
	url     = {https://doi.org/10.1080/0141159031000076110}
}

@article{Mishra2026taas,
  title = {Comparative study of phonon-limited carrier transport in the Weyl semimetal TaAs family},
  author = {Mishra, Shashi B. and Liu, Zhe and Tiwari, Sabyasachi and Giustino, Feliciano and Margine, Elena R.},
  journal = {Phys. Rev. B},
  volume = {113},
  issue = {4},
  pages = {045202},
  numpages = {8},
  year = {2026},
  month = {Jan},
  publisher = {American Physical Society},
  doi = {10.1103/d9np-11j1},
  url     = {https://doi.org/10.1103/d9np-11j1}
}

@article{Liu2025taas,
  title = {Phonon-limited carrier transport in the Weyl semimetal TaAs},
  author = {Liu, Zhe and Mishra, Shashi B. and Lihm, Jae-Mo and Ponc\'e, Samuel and Margine, Elena R.},
  journal = {Phys. Rev. B},
  volume = {112},
  issue = {10},
  pages = {104311},
  numpages = {9},
  year = {2025},
  month = {Sep},
  publisher = {American Physical Society},
  doi = {10.1103/x8zl-w5x3},
  url     = {https://doi.org/10.1103/x8zl-w5x3}
}

@article{loudon1964raman,
title={The Raman effect in crystals},
author={Loudon, Rodney},
journal = {Adv. Phys.},
volume={13},
number={52},
pages={423--482},
year={1964},
publisher={Taylor \& Francis},
	doi     = {10.1080/00018736400101051},
	url     = {https://doi.org/10.1080/00018736400101051}
}

@book{cardona1982light,
	title     = {Light Scattering in Solids II: Basic Concepts and Instrumentation},
	editor    = {Cardona, Manuel and G{\"u}ntherodt, Gernot},
	series    = {Topics in Applied Physics},
	volume    = {50},
	publisher = {Springer},
	address   = {Berlin},
	year      = {1982},
	doi     = {10.1007/3-540-11380-0},
	url     = {https://doi.org/10.1007/3-540-11380-0}
}

@article{perdew1983physical,
title={Physical content of the exact Kohn-Sham orbital energies: band gaps and derivative discontinuities},
author={Perdew, John P and Levy, Mel},
journal = {Phys. Rev. Lett.},
volume={51},
number={20},
pages={1884},
year={1983},
publisher={APS},
	doi     = {10.1103/PhysRevLett.51.1884},
	url     = {https://doi.org/10.1103/PhysRevLett.51.1884}
}

@article{young2012dirac,
title={Dirac semimetal in three dimensions},
author={Young, Steve M and Zaheer, Saad and Teo, Jeffrey CY and Kane, Charles L and Mele, Eugene J and Rappe, Andrew M},
journal = {Phys. Rev. Lett.},
volume={108},
number={14},
pages={140405},
year={2012},
publisher={APS},
	doi     = {10.1103/PhysRevLett.108.140405},
	url     = {https://doi.org/10.1103/PhysRevLett.108.140405}
}

@article{bradlyn2017topological,
title={Topological quantum chemistry},
author={Bradlyn, Barry and Elcoro, Luis and Cano, Jennifer and Vergniory, Maia G and Wang, Zhijun and Felser, Claudia and Aroyo, Mois I and Bernevig, B Andrei},
journal = {Nature},
volume={547},
number={7663},
pages={298--305},
year={2017},
publisher={Nature Publishing Group UK London},
	doi     = {10.1038/nature23268},
	url     = {https://doi.org/10.1038/nature23268}
}

@article{fu2007topological,
title={Topological insulators in three dimensions},
author={Fu, Liang and Kane, Charles L and Mele, Eugene J},
journal = {Phys. Rev. Lett.},
volume={98},
number={10},
pages={106803},
year={2007},
publisher={APS},
	doi     = {10.1103/PhysRevLett.98.106803},
	url     = {https://doi.org/10.1103/PhysRevLett.98.106803}
}

@article{po2017symmetry,
	title   = {Symmetry-based indicators of band topology in the 230 space groups},
	author  = {Po, Hoi Chun and Vishwanath, Ashvin and Watanabe, Haruki},
	journal = {Nat. Commun.},
	volume  = {8},
	number  = {1},
	pages   = {50},
	year    = {2017},
	doi     = {10.1038/s41467-017-00133-2},
	url     = {https://doi.org/10.1038/s41467-017-00133-2}
}

@article{ali2014large,
	title   = {Large, non-saturating magnetoresistance in {WTe$_2$}},
	author  = {Ali, Mazhar N. and Xiong, Jun and Flynn, Steven and Tao, Jing and Gibson, Quinn D. and Schoop, Leslie M. and Liang, Tian and Haldolaarachchige, Neel and Hirschberger, Max and Ong, N. P. and Cava, R. J.},
	journal = {Nature},
	volume  = {514},
	number  = {7521},
	pages   = {205--208},
	year    = {2014},
	doi     = {10.1038/nature13763},
	url     = {https://doi.org/10.1038/nature13763}
}

@article{wang2016origin,
title={Magnetotransport and Fermi surface studies of a purported nodal line semimetal ZrAs 2},
author={Nandi, Suman and Maity, Bishal Baran and Sharma, Vikash and Verma, Rahul and Saini, Vikas and Singh, Bahadur and Aoki, Dai and Thamizhavel, A},
journal = {Phys. Rev. B},
volume={109},
number={7},
pages={075155},
year={2024},
publisher={APS},
	doi     = {10.1103/PhysRevB.109.075155},
	url     = {https://doi.org/10.1103/PhysRevB.109.075155}
}

@article{hasan2010colloquium,
	title   = {Colloquium: Topological insulators},
	author  = {Hasan, M. Z. and Kane, C. L.},
	journal = {Rev. Mod. Phys.},
	volume  = {82},
	number  = {4},
	pages   = {3045--3067},
	year    = {2010},
	doi     = {10.1103/RevModPhys.82.3045},
	url     = {https://doi.org/10.1103/RevModPhys.82.3045}
}

@article{qi2011topological,
  title = {Topological insulators and superconductors},
  author = {Qi, Xiao-Liang and Zhang, Shou-Cheng},
  journal = {Rev. Mod. Phys.},
  volume = {83},
  issue = {4},
  pages = {1057--1110},
  numpages = {0},
  year = {2011},
  month = {Oct},
  publisher = {American Physical Society},
  doi = {10.1103/RevModPhys.83.1057},
  url = {https://link.aps.org/doi/10.1103/RevModPhys.83.1057}
}

@article{armitage2018weyl,
	title   = {Weyl and {Dirac} semimetals in three-dimensional solids},
	author  = {Armitage, N. P. and Mele, E. J. and Vishwanath, Ashvin},
	journal = {Rev. Mod. Phys.},
	volume  = {90},
	number  = {1},
	pages   = {015001},
	year    = {2018},
	doi     = {10.1103/RevModPhys.90.015001},
	url     = {https://doi.org/10.1103/RevModPhys.90.015001}
}

@article{lv2021experimental,
	title   = {Experimental perspective on three-dimensional topological semimetals},
	author  = {Lv, B. Q. and Qian, T. and Ding, H.},
	journal = {Rev. Mod. Phys.},
	volume  = {93},
	number  = {2},
	pages   = {025002},
	year    = {2021},
	doi     = {10.1103/RevModPhys.93.025002},
	url     = {https://doi.org/10.1103/RevModPhys.93.025002}
}

@article{neupane2014observation,
	title   = {Observation of a three-dimensional topological {Dirac} semimetal phase in high-mobility {Cd$_3$As$_2$}},
	author  = {Neupane, Madhab and Xu, Su-Yang and Sankar, R. and Alidoust, N. and Bian, G. and Liu, C. and Belopolski, I. and Chang, T.-R. and Jeng, H.-T. and Lin, H. and Bansil, A. and Chou, F. and Hasan, M. Z.},
	journal = {Nat. Commun.},
	volume  = {5},
	pages   = {3786},
	year    = {2014},
	doi     = {10.1038/ncomms4786},
	url     = {https://doi.org/10.1038/ncomms4786}
}

@article{borisenko2014experimental,
	title   = {Experimental realization of a three-dimensional {Dirac} semimetal},
	author  = {Borisenko, Sergey and Gibson, Quinn and Zhou, Danhui and Evtushinsky, Volodymyr and B{\"u}chner, Bernd and Cava, Robert J.},
	journal = {Phys. Rev. Lett.},
	volume  = {113},
	number  = {2},
	pages   = {027603},
	year    = {2014},
	doi     = {10.1103/PhysRevLett.113.027603},
	url     = {https://doi.org/10.1103/PhysRevLett.113.027603}
}

@article{liu2014discovery,
	title   = {Discovery of a three-dimensional topological {Dirac} semimetal, {Na$_3$Bi}},
	author  = {Liu, Z. K. and Zhou, B. and Zhang, Y. and Wang, Z. J. and Weng, H. M. and Prabhakaran, D. and Mo, S.-K. and Shen, Z. X. and Fang, Z. and Dai, X. and Hussain, Z. and Chen, Y. L.},
	journal = {Science},
	volume  = {343},
	number  = {6173},
	pages   = {864--867},
	year    = {2014},
	doi     = {10.1126/science.1245085},
	url     = {https://doi.org/10.1126/science.1245085}
}

@article{lv2015experimental,
	title   = {Experimental discovery of {Weyl} semimetal {TaAs}},
	author  = {Lv, B. Q. and Weng, H. M. and Fu, B. B. and Wang, X. P. and Miao, H. and Ma, J. and Richard, P. and Huang, X. C. and Zhao, L. X. and Chen, G. F. and Fang, Z. and Dai, X. and Qian, T. and Ding, H.},
	journal = {Phys. Rev. X},
	volume  = {5},
	number  = {3},
	pages   = {031013},
	year    = {2015},
	doi     = {10.1103/PhysRevX.5.031013},
	url     = {https://doi.org/10.1103/PhysRevX.5.031013}
}

@article{xu2015discovery,
	title   = {Discovery of a {Weyl} fermion semimetal and topological {Fermi} arcs},
	author  = {Xu, Su-Yang and Belopolski, Ilya and Alidoust, Nasser and Neupane, Madhab and Bian, Guang and Zhang, Chenglong and Sankar, Raman and Chang, Guoqing and Yuan, Zhujun and Lee, Chi-Cheng and Huang, Shin-Ming and Zheng, Hao and Ma, Jie and Sanchez, Daniel S. and Wang, BaoKai and Bansil, Arun and Chou, Fang-Cheng and Shibayev, Pavel P. and Lin, Hsin and Jia, Shuang and Hasan, M. Zahid},
	journal = {Science},
	volume  = {349},
	number  = {6248},
	pages   = {613--617},
	year    = {2015},
	doi     = {10.1126/science.aaa9297},
	url     = {https://doi.org/10.1126/science.aaa9297}
}

@article{shekhar2015extremely,
	title   = {Extremely large magnetoresistance and ultrahigh mobility in the topological {Weyl} semimetal candidate {NbP}},
	author  = {Shekhar, Chandra and Nayak, Ajit K. and Sun, Yan and Schmidt, Marcus and Nicklas, Michael and Leermakers, Inge and Zeitler, Uli and Skourski, Yurii and Wosnitza, Jochen and Liu, Zhaosheng and Chen, Yulin and Schnelle, Walter and Borrmann, Horst and Grin, Yuri and Felser, Claudia and Yan, Binghai},
	journal = {Nat. Phys.},
	volume  = {11},
	number  = {8},
	pages   = {645--649},
	year    = {2015},
	doi     = {10.1038/nphys3372},
	url     = {https://doi.org/10.1038/nphys3372}
}

@article{xiong2015evidence,
	title   = {Evidence for the chiral anomaly in the {Dirac} semimetal {Na$_3$Bi}},
	author  = {Xiong, Jun and Kushwaha, Satya K. and Liang, Tian and Krizan, Jason W. and Hirschberger, Max and Wang, Wudi and Cava, R. J. and Ong, N. P.},
	journal = {Science},
	volume  = {350},
	number  = {6259},
	pages   = {413--416},
	year    = {2015},
	doi     = {10.1126/science.aac6089},
	url     = {https://doi.org/10.1126/science.aac6089}
}

@article{yang2014dirac,
  title={Classification of stable three-dimensional Dirac semimetals with nontrivial topology},
  author={Yang, Bohm-Jung and Nagaosa, Naoto},
  journal={Nature communications},
  volume={5},
  number={1},
  pages={4898},
  year={2014},
  publisher={Nature Publishing Group UK London}
}

@article{vergniory2019complete,
	title   = {A complete catalogue of high-quality topological materials},
	author  = {Vergniory, M. G. and Elcoro, L. and Felser, Claudia and Regnault, Nicolas and Bernevig, B. Andrei and Wang, Zhijun},
	journal = {Nature},
	volume  = {566},
	number  = {7745},
	pages   = {480--485},
	year    = {2019},
	doi     = {10.1038/s41586-019-0954-4},
	url     = {https://doi.org/10.1038/s41586-019-0954-4}
}

@article{tang2019comprehensive,
	title   = {Comprehensive search for topological materials using symmetry indicators},
	author  = {Tang, Feng and Po, Hoi Chun and Vishwanath, Ashvin and Wan, Xiangang},
	journal = {Nature},
	volume  = {566},
	number  = {7745},
	pages   = {486--489},
	year    = {2019},
	doi     = {10.1038/s41586-019-0937-5},
	url     = {https://doi.org/10.1038/s41586-019-0937-5}
}

@article{zhang2019catalogue,
	title   = {Catalogue of topological electronic materials},
	author  = {Zhang, Tiantian and Jiang, Yi and Song, Zhida and Huang, He and He, Yanying and Fang, Chen and Weng, Hongming and Fang, Zhong and Wang, Zhijun and Chen, Xiao-Qing},
	journal = {Nature},
	volume  = {566},
	number  = {7745},
	pages   = {475--479},
	year    = {2019},
	doi     = {10.1038/s41586-019-0944-6},
	url     = {https://doi.org/10.1038/s41586-019-0944-6}
}

@article{wieder2018wallpaper,
	title   = {Wallpaper fermions and the nonsymmorphic {Dirac} insulator},
	author  = {Wieder, Benjamin J. and Bradlyn, Barry and Wang, Zhijun and Cano, Jennifer and Kim, Youngkuk and Kim, Hyo-Suk and Fradkin, Eduardo and Kane, C. L. and Bernevig, B. Andrei},
	journal = {Science},
	volume  = {361},
	number  = {6399},
	pages   = {246--251},
	year    = {2018},
	doi     = {10.1126/science.aao2800},
	url     = {https://doi.org/10.1126/science.aao2800}
}

@article{hossain2025superconductivity,
	title={Superconductivity and a van Hove singularity confined to the surface of a topological semimetal},
	author={Hossain, Md Shafayat and Islam, Rajibul and Cheng, Zi-Jia and Muhammad, Zahir and Zhang, Qi and Guguchia, Zurab and Krieger, Jonas A and Casas, Brian and Jiang, Yu-Xiao and Litskevich, Maksim and others},
	journal = {Nat. Commun.},
	volume={16},
	number={1},
	pages={3998},
	year={2025},
	publisher={Nature Publishing Group UK London},
	doi     = {10.1038/s41467-025-58024-w},
	url     = {https://doi.org/10.1038/s41467-025-58024-w}
}

@article{muhammad2024electronic,
	title={Electronic transport and interaction of lattice dynamics in topological nodalline semimetal HfAs2 single crystals},
	author={Muhammad, Zahir and Hussain, Ghulam and Islam, Rajbul and Zawadzka, Natalia and Hossain, Md Shafayat and Iqbal, Obaid and Babi{\'n}ski, Adam and Molas, Maciej R and Xue, Fei and Zhang, Yue and others},
	journal = {Adv. Funct. Mater.},
	volume={34},
	number={41},
	pages={2316775},
	year={2024},
	publisher={Wiley Online Library},
	doi     = {10.1002/adfm.202316775},
	url     = {https://doi.org/10.1002/adfm.202316775}
}

@article{huang2015observation,
	title={Observation of the chiral-anomaly-induced negative magnetoresistance in 3D Weyl semimetal TaAs},
	author={Huang, Xiaochun and Zhao, Lingxiao and Long, Yujia and Wang, Peipei and Chen, Dong and Yang, Zhanhai and Liang, Hui and Xue, Mianqi and Weng, Hongming and Fang, Zhong and others},
	journal = {Phys. Rev. X},
	volume={5},
	number={3},
	pages={031023},
	year={2015},
	publisher={APS},
	doi     = {10.1103/PhysRevX.5.031023},
	url     = {https://doi.org/10.1103/PhysRevX.5.031023}
}

@article{naumann2020orbital,
	title={Orbital effect and weak localization in the longitudinal magnetoresistance of Weyl semimetals NbP, NbAs, TaP, and TaAs},
	author={Naumann, M and Arnold, F and Bachmann, MD and Modic, KA and Moll, PJW and S{\"u}{\ss}, V and Schmidt, M and Hassinger, E},
	journal = {Phys. Rev. Mater.},
	volume={4},
	number={3},
	pages={034201},
	year={2020},
	publisher={APS},
	doi     = {10.1103/PhysRevMaterials.4.034201},
	url     = {https://doi.org/10.1103/PhysRevMaterials.4.034201}
}

@article{Giannozzi2017,
author = {Giannozzi, Paolo and Andreussi, Oliviero and Brumme, Thomas and Bunau, Olivier and Buongiorno Nardelli, Marco and Calandra, Matteo and Car, Roberto and Cavazzoni, Carlo and Ceresoli, Davide and Cococcioni, Matteo and Colonna, Nicola and Carnimeo, Ivan and Dal Corso, Andrea and de Gironcoli, Stefano and Delugas, Pietro and DiStasio Jr, Robert A. and Ferretti, Andrea and Floris, Andrea and Fratesi, Guido and Fugallo, Giorgio and Gebauer, Ralph and Gerstmann, Uwe and Giustino, Feliciano and Gorni, Tommaso and Jia, Jia and Kawamura, Masayuki and Ko, Hyo Yoon and Kokalj, Anton and Küçükbenli, Esen and Lazzeri, Michele and Marsili, Matteo and Marzari, Nicola and Mauri, Francesco and Nguyen, Ngoc Linh and Nguyen, Huy-Viet and Otero-de-la-Roza, Alberto and Paulatto, Lorenzo and Poncé, Samuel and Rocca, Dario and Sabatini, Riccardo and Santra, Biswajit and Schlipf, Martin and Seitsonen, Ari P. and Smogunov, Alexander and Timrov, Iurii and Thonhauser, Timo and Umari, Paolo and Vast, Nathalie and Wu, Xifan and Baroni, Stefano},
journal = {J. Phys.: Condens. Matter},
doi = {10.1088/1361-648x/aa8f79},
number = {46},
pages = {465901},
title = {{Advanced capabilities for materials modelling with Quantum ESPRESSO}},
volume = {29},
year = {2017},
	url     = {https://doi.org/10.1088/1361-648x/aa8f79}
}

@article{hamman2013,
  title = {{Optimized norm-conserving Vanderbilt pseudopotentials}},
  author = {Hamann, D. R.},
  journal = {Phys. Rev. B},
  volume = {88},
  issue = {8},
  pages = {085117},
  numpages = {10},
  year = {2013},
  month = {Aug},
  publisher = {American Physical Society},
  doi = {10.1103/PhysRevB.88.085117},
  url     = {https://doi.org/10.1103/PhysRevB.88.085117}
}

@article{Methfessel1989,
  title = {{High-precision sampling for Brillouin-zone integration in metals}},
  author = {Methfessel, M. and Paxton, A. T.},
  journal = {Phys. Rev. B},
  volume = {40},
  issue = {6},
  pages = {3616--3621},
  numpages = {0},
  year = {1989},
  month = {Aug},
  publisher = {American Physical Society},
  doi = {10.1103/PhysRevB.40.3616},
  url     = {https://doi.org/10.1103/PhysRevB.40.3616}
}

@article{gao2021irvsp,
	title   = {{Irvsp}: To obtain irreducible representations of electronic states in the VASP},
	author  = {Gao, Jiacheng and Wu, Quansheng and Persson, Clas and Wang, Zhijun},
	journal = {Comput. Phys. Commun.},
	volume  = {261},
	pages   = {107760},
	year    = {2021},
	doi     = {10.1016/j.cpc.2020.107760},
	url     = {https://doi.org/10.1016/j.cpc.2020.107760}
}

@article{iraola2022irrep,
	title   = {{IrRep}: Symmetry eigenvalues and irreducible representations of ab initio band structures},
	author  = {Iraola, Mikel and Ma{\~n}es, Juan L. and Bradlyn, Barry and Horton, Matthew K. and Neupert, Titus and Vergniory, Maia G. and Tsirkin, Stepan S.},
	journal = {Comput. Phys. Commun.},
	volume  = {272},
	pages   = {108226},
	year    = {2022},
	doi     = {10.1016/j.cpc.2021.108226},
	url     = {https://doi.org/10.1016/j.cpc.2021.108226}
}

@article{hung2024qeraman,
	title   = {{QERaman}: An open-source program for calculating resonance Raman spectra based on {Quantum ESPRESSO}},
	author  = {Hung, Nguyen T. and Huang, Jianqi and Tatsumi, Yuki and Yang, Teng and Saito, Riichiro},
	journal = {Comput. Phys. Commun.},
	volume  = {295},
	pages   = {108967},
	year    = {2024},
	doi     = {10.1016/j.cpc.2023.108967},
	url     = {https://doi.org/10.1016/j.cpc.2023.108967}
}

@article{Pizzi2020,
	doi = {10.1088/1361-648x/ab51ff},
	url     = {https://doi.org/10.1088/1361-648x/ab51ff},
	year = 2020,
	month = {jan},
	publisher = {{IOP} Publishing},
	volume = {32},
	number = {16},
	pages = {165902},
  author = {Pizzi, Giovanni and Vitale, Valerio and Arita, Ryotaro and Bl{\"u}gel, Stefan and Freimuth, Frank and G{\'e}ranton, Guillaume and Gibertini, Marco and Gresch, Dominik and Johnson, Charles and Koretsune, Takashi and Iba{\~n}ez-Azpiroz, Julen and Lee, Hyungjun and Lihm, Jae-Mo and Marchand, Daniel and Marrazzo, Antimo and Mokrousov, Yuriy and Mustafa, Jamal I. and Nohara, Yoshiro and Nomura, Yusuke and Paulatto, Lorenzo and Ponc{\'e}, Samuel and Ponweiser, Thomas and Qiao, Junfeng and Th{\"o}le, Florian and Tsirkin, Stepan S. and Wierzbowska, Ma{\l}gorzata and Marzari, Nicola and Vanderbilt, David and Souza, Ivo and Mostofi, Arash A. and Yates, Jonathan R.},
	title = {{Wannier90 as a community code: new features and applications}},
	journal = {J. Phys.: Condens. Matter}
}

@article{Perdew1996,
  title = {{Generalized Gradient Approximation Made Simple}},
  author = {Perdew, John P. and Burke, Kieron and Ernzerhof, Matthias},
  journal = {Phys. Rev. Lett.},
  volume = {77},
  issue = {18},
  pages = {3865--3868},
  numpages = {0},
  year = {1996},
  month = {Oct},
  publisher = {American Physical Society},
  doi = {10.1103/PhysRevLett.77.3865},
  url     = {https://doi.org/10.1103/PhysRevLett.77.3865}
}

@article{kawamura2014tetrahedron,
	title   = {Improved tetrahedron method for the Brillouin-zone integration applicable to response functions},
	author  = {Kawamura, Mitsuaki and Gohda, Yoshihiro and Tsuneyuki, Shinji},
	journal = {Phys. Rev. B},
	volume  = {89},
	number  = {9},
	pages   = {094515},
	year    = {2014},
	publisher = {APS},
	doi     = {10.1103/PhysRevB.89.094515},
	url     = {https://doi.org/10.1103/PhysRevB.89.094515}
}

@article{muhammad2026chiralphonon,
      title={{Chiral Phonons and Giant Anisotropic Photoresponse in Quasi-1D van der Waals Semiconductor ZrSnS3}}, 
      author={Zahir Muhammad and Shashi B. Mishra and Gayatri and Grzegorz Krasucki and Wajid Ali and Katarzyna Olkowska-Pucko and Obaid Iqbal and Zia Ur Rehman and Aziz Ur Rahman and Maciej R. Molas and Lin Xiaoyang and Weisheng Zhao},
      year={2026},
      journal={arXiv:2608.22700},
      url={https://arxiv.org/abs/2608.22700}, 
}

@article{Lifshitz1960,
  title = {Anomalies of electron characteristics of a metal in the high pressure region},
  author = {Lifshitz, I. M.},
  journal = {Sov. Phys. JETP},
  volume = {11},
  pages = {1130--1135},
  year = {1960}
}

@article{Sobota2021RMP,
  title = {Angle-resolved photoemission studies of quantum materials},
  author = {Sobota, Jonathan A. and He, Yu and Shen, Zhi-Xun},
  journal = {Rev. Mod. Phys.},
  volume = {93},
  issue = {2},
  pages = {025006},
  year = {2021},
  doi = {10.1103/RevModPhys.93.025006}
}

@article{Damascelli2003RMP,
  title = {Angle-resolved photoemission studies of the cuprate superconductors},
  author = {Damascelli, Andrea and Hussain, Zahid and Shen, Zhi-Xun},
  journal = {Rev. Mod. Phys.},
  volume = {75},
  issue = {2},
  pages = {473--541},
  year = {2003},
  doi = {10.1103/RevModPhys.75.473}
}

\end{document}